\documentclass{emulateapj}
\usepackage{epsfig,pstricks,graphicx}

\slugcomment{Accepted for publication in the Astrophysical Journal}
\shortauthors{Tripp and Song}
\shorttitle{Outer Galaxy High-Velocity Clouds}

\begin{document}

\title{The 21cm ``Outer Arm'' and the Outer-Galaxy High-Velocity Clouds: Connected By Kinematics, Metallicity, and Distance\altaffilmark{1}}

\altaffiltext{1}{Based on observations with (1) the NASA/ESA {\it
    Hubble Space Telescope}, obtained at the Space Telescope Science
  Institute, which is operated by the Association of Universities for
  Research in Astronomy, Inc., under NASA contract NAS 5-26555, and
  (2) the NASA-CNES/ESA {\it Far Ultraoviolet Spectroscopic Explorer}
  mission, operated by Johns Hopkins University, supported by NASA
  contract NAS 5-32985.}

\author{Todd M. Tripp and Limin Song}

\affil{Department of Astronomy, University of Massachusetts,
  Amherst, MA 01003}
\email{tripp@astro.umass.edu}

\begin{abstract}
Using high-resolution ultraviolet spectra obtained with the
\textit{HST} Space Telescope Imaging Spectrograph (STIS) and the
\textit{Far Ultraviolet Spectroscopic Explorer}, we study the
metallicity, kinematics, and distance of the gaseous ``Outer Arm''
(OA) and the high-velocity clouds (HVCs) in the outer Galaxy. We
detect the OA in a variety of absorption lines toward two QSOs,
H1821+643 and HS0624+6907.  We search for OA absorption toward eight
Galactic stars and detect it in one case, which constrains the OA
Galactocentric radius to 9$<R_{G}<$18 kpc. We also detect HVC Complex
G, which is projected near the OA at a similar velocity, in absorption
toward two stars; Complex G is therefore in the same region at
$R_{G} = 8 - 10$ kpc.  HVC Complex C is known to be at a similar
Galactocentric radius. Toward H1821+643, the low-ionization absorption
lines are composed of multiple narrow components, indicating the
presence of several cold clouds and rapid cooling and fragmentation.
Some of the highly ionized gas is also surprisingly cool.  Accounting
for ionization corrections, we find that the OA metallicity is
$Z=0.2-0.5 Z_{\odot}$, but nitrogen is underabundant and some species
are possibly mildly depleted by dust. The similarity of the OA
metallicity, Galactocentric location, and kinematics to those of the
adjacent outer-Galaxy HVCs, including high velocities that are not
consistent with Galactic rotation, suggests that the OA and
outer-Galaxy HVCs could have a common origin.
\end{abstract}

\keywords{Galaxies:ISM --- ISM: abundances --- quasars: absorption lines --- quasars: individual (H1821+643, HS0624+6907)}

\section{Introduction}
Absorption-line measurements of metal abundances, physical conditions,
and kinematics of diffuse gas in well-constrained locations of the
outer Milky Way (e.g., Savage et al. 1995) are relatively rare,
particularly near the plane of the Galaxy (Wakker 2001).  Such
measurements can provide unique insights into galaxy evolution for
several related reasons:

First, the chemical enrichment patterns at large Galactocentric
distances, and the corresponding implications regarding the Galactic
abundance gradient, are fundamental observables that can be compared
to theory in order to understand a variety of processes that affect
the evolution of a galaxy (e.g., Chiappini et al. 2001; Carigi et
al. 2005). For example, the processes by which galaxies acquire gas
and fuel new star formation are not entirely understood, and this has
become a key question in current galaxy evolution studies.  Recent
theoretical work has indicated that gas accretion might not occur
through a spherical accretion shock as traditionally envisioned but
instead could occur in a ``cold mode'' in which the matter flows into
galaxies in filaments that radiatively cool and never approach the
virial temperature (Katz et al. 2003; Kere\v{s} et al. 2005,2009;
Birnboim \& Dekel 2003).  To test models of gas accretion, it is
necessary to measure the properties of incoming matter in various
locations.  Some of the models indicate that accreting matter can have
substantial angular momentum and could settle into the outer disk
(e.g., Kere\v{s} et al. 2005), a hypothesis that has some support from
recent observations (Moran et al. 2012).  In this case, it is also
important to probe how the incoming gas is transported into the inner
disk to enable new star formation or even central black hole growth
and activity (Hopkins \& Quataert 2010).  Conversely, star formation
can generate galactic fountain {\it outflows} that remove enriched
matter from the inner galaxy.  Some of outflowing material could
escape entirely, but it is also probable that some of this enriched
matter will return to the outer disk and add to the matter reservoir
for subsequent star formation (e.g., Bregman 1980; Oppenheimer et
al. 2010; Lehner \& Howk 2011).  These competing inflow and outflow
processes should have significantly different abundance signatures:
inflowing intergalactic matter should have relatively low metallicity
while the outflowing fountain material is expected to be metal
enriched, and indeed, low-redshift observations have shown that halo
gas can be remarkably metal-poor (e.g., Tripp et al. 2005; Ribaudo et
al. 2011) as well as highly enriched (e.g., Jenkins et al. 2005;
Prochaska \& Hennawi 2009; Tripp et al. 2011).  Thus, these flows
should have opposite effects on galactic abundance gradients, and
ultimately a complete galactic evolution model should be able to
explain observed metallicities in both the inner and outer regions of
a galaxy.  Similarly, it is important to constrain the microphysics of
inflowing/outflowing matter -- how (and where) are the flows ionized,
ablated, and mixed into the general ISM?  How do they subsequently
cool to reach conditions suitable for new star formation?  Some
galactic wind models drive the outflows by radiative pressure and may
require significant quantities of dust in the outflowing material
(Aguirre et al. 2001; Murray et al. 2011).  Absorption-line
measurements can reveal the presence of dust in the outer galaxy
through abundance patterns indicative of depletion onto dust grains
(e.g., Savage \& Sembach 1996; Jenkins 2009), so outer-galaxy
absorption studies can also shed insight on the roles played by dust
in inflows and outflows.

Second, absorption-line observations of damped Ly$\alpha$ absorbers
[DLAs; QSO absorbers with $N$(\ion{H}{1}) $> 2 \times 10^{20}$
  cm$^{-2}$) and sub-DLAs ($1 \times 10^{19} \lesssim N$(\ion{H}{1})
  $\leq 2 \times 10^{20}$] provide sensitive probes of the chemical
enrichment history of the Universe from $z = 0$ to $z > 4$ (e.g.,
Prochaska et al. 2003; Wolfe et al. 2005, Meiring et al. 2009,2011),
but the context of the DLAs (i.e., the environment and nature of the
absorbers) is generally hard to study, and currently only limited
information is available regarding the origins of DLAs (e.g., Chen \&
Lanzetta 2003; Rao et al. 2003; Battisti et al. 2012). Considering the
cross section of a disk or halo of gas, it is likely that many of the
DLAs arise in the outer regions of galaxies. The Milky Way is a damped
Ly$\alpha$ absorber. Since its absorption context can be scrutinized
in great detail, the Milky Way provides a valuable laboratory for
understanding the nature of DLAs/sub-DLAs, but measurements of
abundances patterns in the more distant {\it outer} Galaxy are still
relatively limited.

Third, while some abundances in the outer galaxy and abundance
gradients have been measured using \ion{H}{2} region emission lines,
there has long been concern about whether \ion{H}{2} region abundances
are biased by ``self pollution'', i.e. whether the abundances are
boosted by freshly formed metals from massive stars embedded within
the \ion{H}{2} region.  Abundances from absorption lines in distant
background objects probe random locations with respect to foreground
\ion{H}{2} regions and can test the self-pollution hypothesis, and
some absorption-line studies have indeed suggested that self pollution
does occur (Cannon et al. 2005 and references therein).  The discovery
that many gas-rich galaxies have ``extended ultraviolet disks''
(Thilker et al. 2007) underscores the importance of investigating this
issue -- \ion{H}{2} regions may be biased in favor of the UV-bright
star clusters that comprise the extended UV disks and might not
accurately represent abundance gradients for testing theoretical work
as discussed above.

Fourth, outer galaxy abundances can be used to investigate whether
some gaseous structures of the Milky Way could be due to interactions
with satellite galaxies.  This is related to the questions raised
above: one means to bring gas into galaxies is to strip the
interstellar media of dwarf satellite galaxies as they plunge into,
and merge with, the central galaxy.  Satellite interactions can also
stimulate the growth of galactic structures such as warps (e.g.,
Weinberg \& Blitz 2006; Quillen et al. 2009) and may drive continuing
(low-level) star formation in elliptical galaxies (Kaviraj et
al. 2011).  Since satellite galaxies can have significantly different
abundances compared to each other and the Galactic disk, the outer
galaxy abundances provide clues about the origins of galactic
structures and the importance of this mechanism for bringing
additional gas into galaxies.

For these reasons we have conducted a study of the abundances in the
outer galaxy using absorption lines recorded in high-resolution
spectra of two QSOs observed with the {\it Hubble Space Telescope
  (HST)} and the {\it Far Ultraviolet Spectroscopic Explorer
  (FUSE)}. We selected two QSOs for this study, HS0624+6907 and
H1821+643.  These sight lines are unique among the QSOs and AGNs that
have been in observed in the ultraviolet at high spectral resolution
and with good signal-to-noise (S/N) ratios because the QSOs are at
relatively low Galactic latitudes and thus provide an opportunity to
study outer-galaxy gas near the plane.  Moreover, these QSOs lie
behind a high-velocity gaseous structure with a large angular extent
in the outer galaxy known as the ``Outer Arm'' (OA).  This structure
is also unique because its distance has recently been constrained,
which is valuable for understanding its nature and implications.  In
addition, there are several high-velocity cloud complexes near the OA
with similar kinematics, including Complexes C, G, and H.  The {\it
  HST} archive includes high-quality spectra of many stars in the
directions of these gas clouds, and these stellar spectra provide an
opportunity to obtain new constraints on the distance of these
objects.  In this paper, we present a study of these outer-Galaxy
gaseous structures.  We focus on the abundances and physical
conditions in the Outer Arm and whether the OA is related to the HVCs
in its proximity.  In \S~\ref{outerarmnotes}, we provide some comments
on the Outer Arm and the QSO sight lines that probe this part of the
Galaxy. We present the QSO and star observations and absorption-line
measurements in \S~\ref{obssec} and \S~\ref{absmeas}, respectively,
and we discuss new constraints on the distance of the Outer Arm, and
the nearby high-velocity cloud Complex G, in
\S~\ref{distance_section}. In \S~\ref{ionabun}, we examine the
physical conditions of the OA, and we use models to evaluate the
impact of ionization corrections on the metallicity measurements.  We
also make some remarks on the nature of the highly ionized gas in the
Outer Arm.  We discuss our results in \S~\ref{disc} with an emphasis
on the possible origin and implications of the Outer Arm.

\section{QSO Sight Lines Through the Outer Arm}
\label{outerarmnotes}

\begin{figure*}
\epsscale{1.0}
\plotone{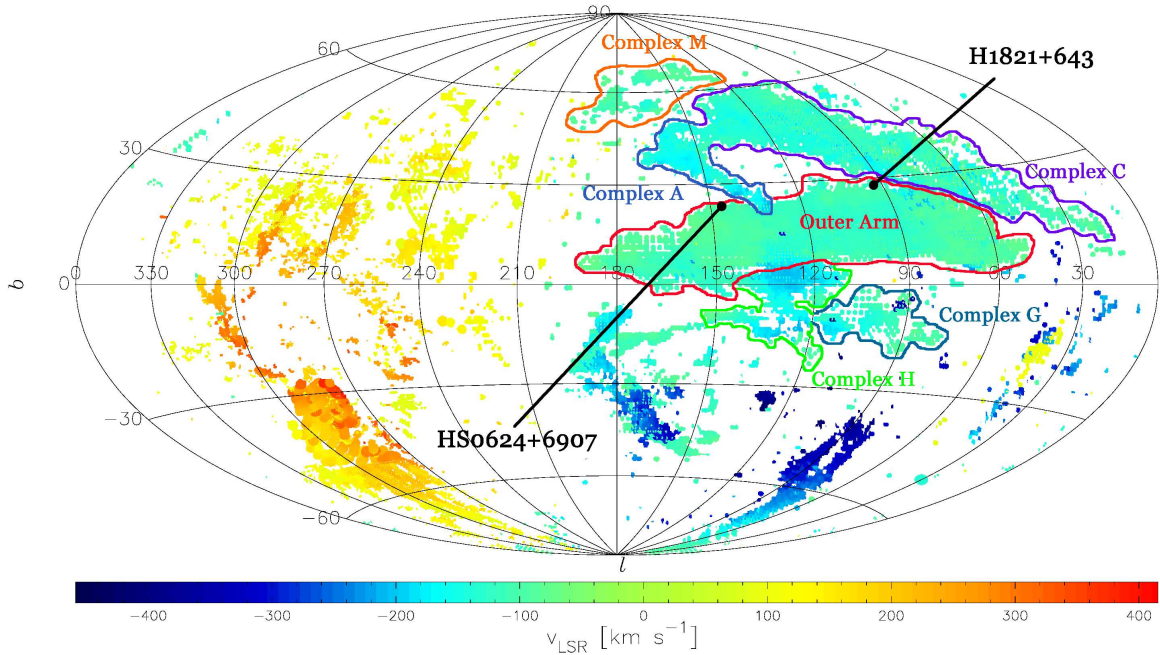}
\caption{Locations of the QSO sight lines H1821+643 and HS0624+6907
  with respect to the 21 cm emission from the Outer Arm.  This map,
  from Wakker et al. (2003), shows 21 cm emission from the Galactic
  high-velocity clouds (HVCs) and the Outer Arm in Galactic
  coordinates (longitude increases from right to left along the
  $x$-axis). LSR velocities are indicated by color using the scale at
  the bottom of the figure, and several of the HVC complexes near the
  Outer Arm are labeled.  \label{outerarm}}
\end{figure*}

The Outer Arm is a large, contiguous neutral hydrogen complex located
in Galactic coordinates at $49 \degr \lesssim l \lesssim 180 \degr$
and $4\degr \lesssim b \lesssim 31 \degr $ over a velocity range $-150
\lesssim v_{\rm LSR} \lesssim -100$ km s$^{-1}$.
Figure~\ref{outerarm} shows an all-sky map of high-velocity 21 cm
emission from Wakker et al. (2003) including the Outer Arm.  The
structure has long been known from 21cm emission studies (e.g.,
Westerhout 1957; Kepner 1970; Burton \& te Lintel Hekkert 1986), and
it is often interpreted to be gas in the outermost Galactic spiral arm
and/or the warped region of the outer disk at a Galactocentric
distance $R_{\rm G} \approx$ 15 kpc (e.g., Kepner 1970; Diplas \&
Savage 1991; Haud 1992). However, whether this gaseous structure is
really a spiral arm has long been questioned (e.g., Davies 1972;
Weaver 1974), and other hypotheses regarding the nature of this object
remain viable. The OA extends to substantial latitudes
(Figure~\ref{outerarm}), which implies a large $z-$height that may
difficult to reconcile with an origin in the Galactic warp, and in
some regions it exhibits large deviations from Galactic rotation
speeds (e.g., Wakker 2001) that are problematic for the spiral arm
interpretation.  Indeed, recently obtained contraints on the OA
distance and kinematics (Lehner \& Howk 2010) call into question
whether this gas cloud is indeed related to the Galactic warp and
outer spiral arm (see below).  Moreover, the similarity of the
kinematics and distance of the OA and the nearby HVCs Complex C and
Complex G (see \S \ref{distance_section}) suggests that the OA and
Complexes C and G (and possibly H) could be related.  Such a complex
configuration of HVCs would not be expected in the warp/spiral arm
interpretation.

The 21 cm emission from the Outer Arm is detected at velocities
similar to expected velocities for a rotating Galactic disk in this
general direction, and consequently, the OA is often ignored in
studies of Galactic HVCs, although it is included in the HVC
compendium of Wakker (2001).  Recently, Lehner \& Howk (2010) have
used the Cosmic Origins Spectrograph (COS) on {\it HST} to detect the
OA in ultraviolet absorption lines toward HS1914+7139 ($l = 103\degr,
b = +24 \degr$), a B2.5 IV star at an estimated Heliocentric distance
of 14.9 kpc.  Interestingly, {\it they detect OA absorption toward
  this star at two velocities}, $v_{\rm LSR} = -118$ and $-180$ km
s$^{-1}$.  This observation places an upper limit\footnote{The
  Galactocentric radius originally reported in Lehner \& Howk (2010)
  was miscalculated (N. Lehner, private communication), and the
  correct $R_{\rm G}$ is listed here.}  on the distance to the OA: the
Galactocentric radius $R_{\rm G}({\rm OA}) < 17.6$ kpc. This is in
agreement with the distance typically derived for the OA, assuming it
is part of the Galactic warp, i.e., $R_{\rm G} \approx$ 15 kpc (Haud
1992).  The OA component at $v_{\rm LSR} = -118$ km s$^{-1}$ is
roughly consistent with corotation at the usually adopted OA distance,
although this velocity implies that the OA should be at a
Galactocentric radius that is $\approx 5$ kpc farther out than the
$R_{\rm G}$ upper limit.  The other velocity component detected by
Lehner \& Howk, $v_{\rm LSR} = -180$ km s$^{-1}$, is more interesting:
this velocity is highly inconsistent with a rotating-disk origin;
corotating disk gas would be far beyond the star at this velocity,
even with the increased Milky Way rotation speed derived by Reid et
al. (2009).  A small portion of the OA shows 21cm emission at
velocities near $-180$ km s$^{-1}$ (see Figure 1 in Tripp et
al. 2003), but mostly the OA is not detected at this velocity in 21 cm
emission.  However, the HVC Complex C, which is close to the OA as
shown in Figure~\ref{outerarm}, has a pervasive and highly extended
``high-velocity ridge'' at $v \approx -180$ km s$^{-1}$ (Tripp et
al. 2003).  This high-velocity ridge has a similar morphology to the
lower-velocity part of Complex C and is almost certainly a component
of Complex C.  The discovery by Lehner \& Howk that the Outer Arm has
a similar higher-velocity component at $v \approx -180$ km s$^{-1}$
suggests that the OA and Complex C could be closely related.  That the
higher-velocity part of the OA is not apparent in most of the 21cm
emission map likely indicates that the \ion{H}{1} column density in
the $v_{\rm LSR} = -180$ km s$^{-1}$ component is too low to be
detected in 21cm emission. The $v_{\rm LSR} = -180$ km s$^{-1}$
component is significantly weaker in the UV absorption lines than the
$v_{\rm LSR} = -118$ km s$^{-1}$ component (see Figure 1 in Lehner \&
Howk 2010), which is consistent with its absence in the 21cm emission
map.  All together, these observations suggest that Complex C and the
OA (and possibly some of the other nearby HVCs) could have a common
origin, and the gas becomes more highly ionized and ablated
approaching the plane, an idea which is supported by the
\ion{O}{6}/\ion{H}{1} ratios in Complex C (Tripp et al. 2003; Sembach
et al. 2003) as well as its morphology in high-resolution maps (Hsu et
al. 2011).  However, other explanations remain viable.  For example,
the HVCs with higher absolute velocities tend to be more highly
ionized, and it is possible that as the HVCs approach the disk, they
are decelerated and compressed, and the increased density from
compression moves the gas into a less ionized state.

\begin{figure*}
\plotone{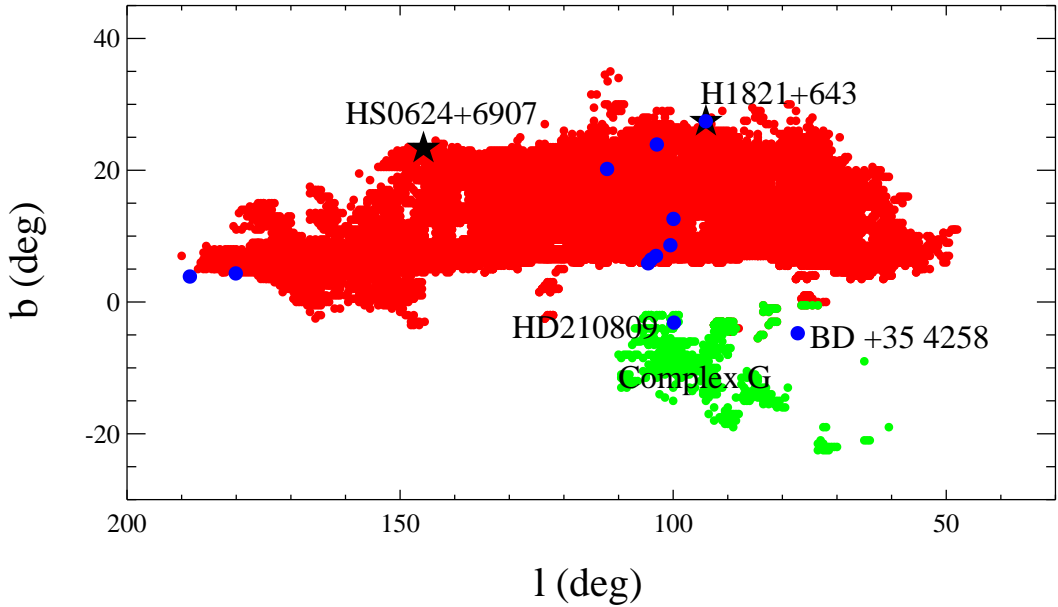} 
\epsscale{1.0} 
\caption{Map of the 21cm emission from the Outer Arm (red dots) and
  Complex G (green dots) from the Leiden-Agentine-Bonn Survey
  (Kalberla \& Haud 2006).  The locations of stars that we have used
  to search for absorption at Outer Arm velocities (see
  Table~\ref{distance_tab}) are marked with blue circles, and the
  sight lines to the QSOs H1821+643 and HS0624+6907 are plotted with
  black stars. Note that the sight line to the central star of the
  planetary nebula K1-16 is very close to the H1821+643 sight line and
  falls nearly on top of the H1821+643 point. Toward BD +35 4258 and
  HD210809, highly significant UV absorption lines are detected at the
  velocity of Complex G (see Figures~\ref{fig_bd35_compG} -
  \ref{fig_hd210809_complexg}). \label{zoommap}}
\end{figure*}

\begin{figure}
\epsscale{1.3}
\plotone{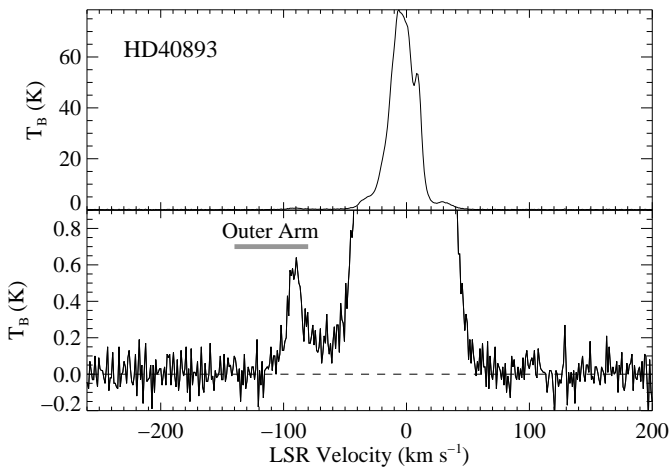}
\caption{H~\textsc{i} 21 cm emission in the direction of HD40893 from the
  Leiden/Argentine/Bonn (LAB) survey (Kalberla et al. 2005). Both
  panels plot brightness temperature vs. LSR velocity.  In the lower
  panel, the gray bar shows the velocity range over which the Outer
  Arm is detected in absorption toward HS0624+6907.\label{highl_1}}
\end{figure}

\begin{figure}
\epsscale{1.3}
\plotone{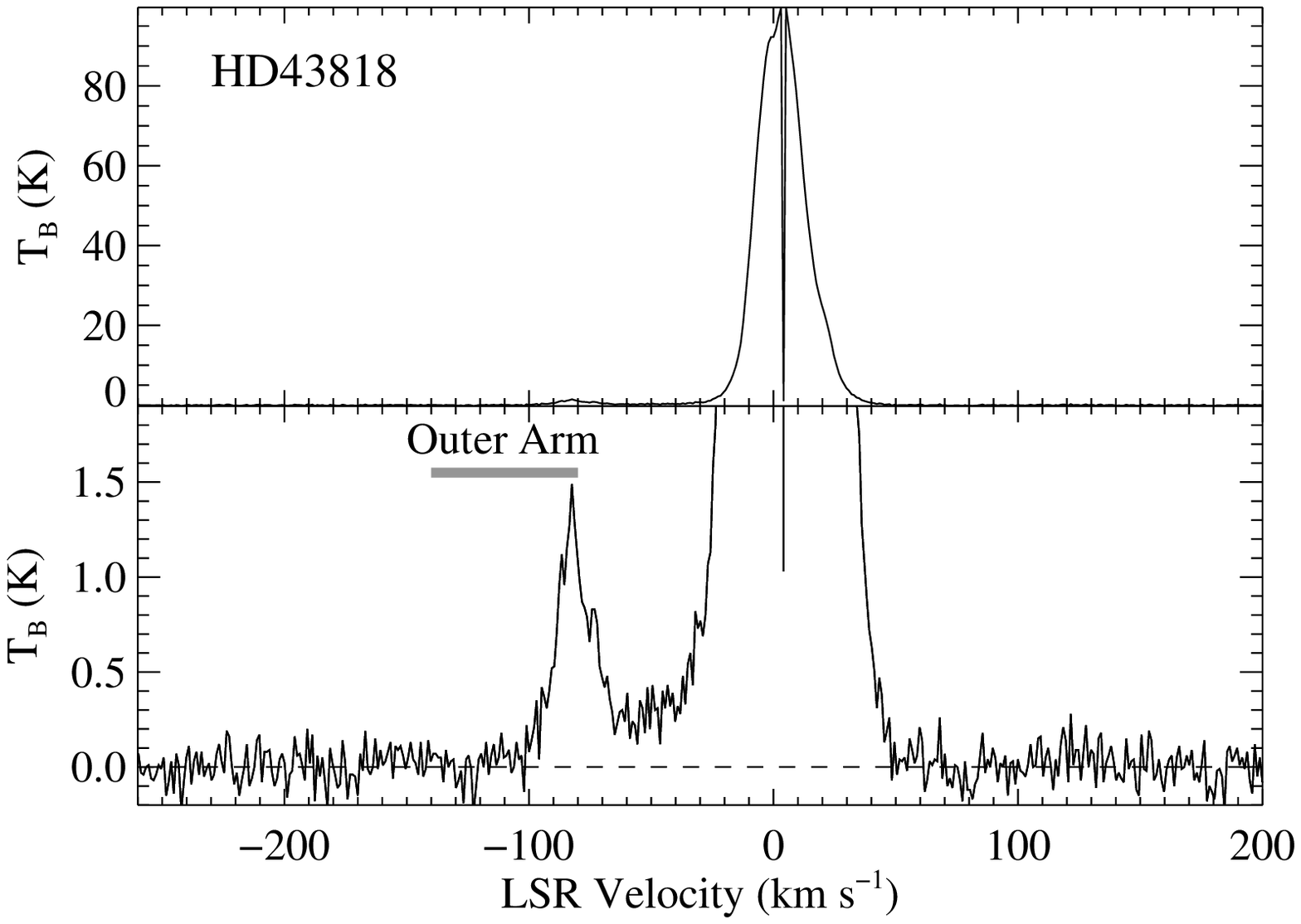}
\caption{H~\textsc{i} 21 cm emission in the direction of HD43818 from the
  LAB Survey, as in Figure~\ref{highl_1}. \label{highl_2}}
\end{figure}

The highly anomalous velocity observed by Lehner \& Howk indicates
that the OA is truly a high-velocity cloud.  If the OA is an outer
spiral arm or part of the Galactic warp, then a mechanism to produce
the high-velocity ($-180$ km s$^{-1}$) component must be identified.
Moreover, the striking similarity of the Outer Arm and Complex C, both
at the ``main'' HVC velocity and in the high-velocity ridge, must be
explained.  If Complex C and the Outer Arm are really related, then
the OA might not be a spiral arm, and it is worthwhile to investigate
other possible origins for this large \ion{H}{1} cloud.  We note that
in this case, the ``Outer Arm'' shown in Figure~\ref{outerarm} may be
a different entity, at least in part, than the Outer Arm on the far
side of the galaxy (e.g., Strasser et al. 2007; Dame \& Thaddeus
2011), which is observed at lower latitudes and lower velocities.

Ultraviolet absorption lines provide a powerful means to learn about
the nature of the Outer Arm. To probe the nature of the OA using UV
absorption spectroscopy, we present below a study of two bright
low-redshift QSOs, H1821+643 ($l=94.00,^{\circ} b=27.42,^{\circ}
z_{\rm QSO} = 0.297$) and HS0624+6907 ($l=145.71,^{\circ}
b=23.35,^{\circ} z_{\rm QSO} = 0.370$), lying in Galactic directions
that pierce the Outer Arm (Figure~\ref{outerarm}). The sight line to
the bright, low-redshift quasar H1821+643 ($z=0.297,V=14.24$) has been
often used in studies of the galactic and intergalactic medium (e.g.,
Savage et al. 1995,1998; Tripp et al. 2000,2001; Oegerle et al. 2000;
Narayanan et al. 2010), and consequently there are high-quality,
high-resolution UV spectra of this target available from the {\it HST}
and {\it FUSE} archives. HS0624+6907 was included in the survey of
Tripp et al. (2008), and thus high-quality UV spectra are also
available for this sight line.

A number of stars in the general direction of the Outer Arm (and its
adjacent HVCs) have also been observed with high resolution and high
S/N with the Space Telescope Imaging Spectrograph (STIS) on {\it HST}.
We will examine the STIS spectra of these stars to bolster the OA
distance constraints from Lehner \& Howk (2010).  For reference, we
overplot the locations of the targets on the 21 cm emission map of the
Outer Arm from the Leiden/Argentine/Bonn (LAB) survey (Kalberla et
al. 2005; Kalberla \& Haud 2006) in Figure~\ref{zoommap}.  We will
show below that the HVC Complex G, which has similar kinematics to
Complexes C and H and the Outer Arm (see Figure~\ref{outerarm}), is at
a similar distance as the OA and Complex C.  To aid this discussion,
we also show the location of Complex G in Figure~\ref{zoommap} along
with the locations of two stars that constrain its distance.  From
Figure~\ref{zoommap}, it is difficult to judge if the two stellar
sight lines (HD40893 and HD43818) at the high-longitude tip of the OA
are truly in the direction of this cloud.  However, as shown in
Figures~\ref{highl_1} and \ref{highl_2}, the LAB spectra in the
directions of these stars clearly show 21 cm emission at the OA
velocity, and we will show below that UV absorption is detected at OA
velocities in the spectrum of HD 43818 (\S \ref{distance_section}).
We note that HD43818 is in a direction where Milky Way gas is known to
exhibit kinematic complexity (e.g., Burton \& Moore 1979) including
the ``anticenter shell'' (e.g., Heiles 1984).  It has been argued that
this structure is not a coherent shell (e.g., Tamanaha 1997), and we
agree: the positive-latitude portion of the anticenter shell connects
contigously in space and velocity with the Outer Arm, and the
negative-latitude part is a component of a large and distinct complex
of HVCs with a much larger angular extent than the original anticenter
shell (see Figure 1).

\section{Observations}
\label{obssec}

\begin{deluxetable*}{llcll}
\tablecaption{Log of STIS Echelle Observations of Stars in the Direction of the Outer Arm\label{star_obs_log}}
\tablehead{Star & \ Observation & Integration & STIS & MAST ID\tablenotemark{b}  \\
            \   & \ \ \ \ \ \ Date        & Time & Grating\tablenotemark{a} & \ \\
            \   &  \            & (seconds) & \ & \ }
\startdata
HD40893\dotfill  & 2004 Feb. 20 & 10064 & E140H & O8NA02010,20 \\
HD43818\dotfill  & 2001 Apr. 9  & 1440 & E140M & O5C07I010 \\
HD198781\dotfill & 1999 Sept. 5 & 360 & E140H & O5C049010 \\ 
HD201908\dotfill & 1999 Sept. 21 & 360 & E140H & O5C051010 \\ 
HD203374\dotfill & 2002 Dec. 24 & 600 & E140M & O6LZ90010 \\
HD207198\dotfill & 2000 Oct. 30 & 4711 & E140H & O59S06010,20 \\
HD208440\dotfill & 1999 Nov. 5  & 720 & E140H & O5C06M010 \\ 
HD209339\dotfill & 2002 Aug. 7  & 1200 & E140H & O6LZ92010 \\
     \           & 2006 Dec. 28 & 1416 & E140H & O5LH0B010,20 \\
HD210809\dotfill & 1999 Oct. 30 & 720 & E140H & O5C01V010 \\
BD+35 4258\dotfill & 2003 Mar. 15 & 1200 & E140M & O6LZ89010 \\
\enddata
\tablenotetext{a}{All of the observations used the $0.2'' \times 0.2''$ STIS aperture except the observation of HD207198, which was recorded with the $0.2'' \times 0.09''$ slit, and the longer-wavelength observation of HD209339 (obtained 2006 Dec. 28), which used the $0.1'' \times 0.03''$ aperture.}
\tablenotetext{b}{Identification code for the data in the Multimission Archive at Space Telescope (see http://archive.stsci.edu/index.html).}
\end{deluxetable*}

\begin{deluxetable*}{lcccc}
\tablewidth{0pc}
\tablecaption{Outer-Arm Profile-Fitting Measurements: H1821+643\label{tab:h1821}}
\tablehead{Species \ \ \ & Fitted Lines & $v$ (LSR) & b & log [$N$ (cm$^{-2}$)] \\
             \     & (\AA )       & (km s$^{-1}$) & (km s$^{-1}$) & \ }
\startdata
O~I\dotfill        & 971.73, 976.45, 1302.17 & $-145\pm 1$ & $4\pm 1$ & 14.43$\pm$0.15 \\
   \               &     \                   & $-131\pm 1$ & $7\pm 2$ & 14.64$\pm$0.11 \\
   \               &     \                   & $-111\pm 1$ & $9\pm 3$ & 14.06$\pm$0.09 \\
N~I\dotfill        & 1199.55\tablenotemark{a}                 & $-146\pm 2$: & $6^{+4}_{-3}$: & 13.19$\pm$0.13 \\
   \               &                         & $-133\pm 1$: & $2^{+4}_{-1}$: & 13.14$\pm$0.19 \\
S~II\dotfill       & 1253.81,1259.52\tablenotemark{b}         & $-131\pm 2$: & $15\pm 1$: & 14.02$\pm$0.08: \\
Al~II\dotfill      & 1670.79                 & $-148\pm 2$ & $4^{+8}_{-3}$ & 12.12$\pm$0.35 \\ 
    \              &     \                   & $-132\pm 4$ & $12^{+8}_{-5}$ & 12.82$\pm$0.26 \\
    \              &     \                   & $-113\pm 8$ & $11^{+9}_{-5}$ & 12.45$\pm$0.58 \\
Si~II\dotfill      & 1190.42, 1193.29, 1260.42, & $-147\pm 1$ & $5\pm 1$ & 13.27$\pm$0.20 \\ 
   \               & 1304.37, 1526.71           & $-134\pm 1$ & $9\pm 3$ & 13.88$\pm$0.16 \\
   \               &     \                      & $-117\pm 4$ & $11\pm 4$ & 13.71$\pm$0.19 \\
Fe~II\dotfill      & 1121.98, 1125.45, 1142.37 & $-145\pm 1$  & $4^{+11}_{-3}$ & 13.15$\pm$0.55 \\  
   \               & 1143.23, 1144.94, 1608.45 & $-134\pm 1$  & $3^{+4}_{-2}$ & 13.25$\pm$0.31 \\
   \               &   \                       & $-128\pm 9$  & $15^{+9}_{-6}$ & 13.64$\pm$0.31 \\
C~IV\dotfill       & 1548.20,1550.78         & $-213\pm 1$ & $9\pm 1$ & 13.25$\pm$0.03 \\
   \               &     \                   & $-151\pm 3$ & $5^{+10}_{-3}$ & 12.59$\pm$0.37 \\
   \               &     \                   & $-131\pm 4$ & $14^{+8}_{-5}$ & 13.28$\pm$0.40 \\
   \               &     \                   & $-111\pm 14$ & $33\pm 11$ & 13.80$\pm$0.09 \\
Si~IV\dotfill      & 1393.76,1402.78\tablenotemark{c}         & $-150\pm 1$  & $4\pm 1$ & 12.40$\pm$0.11 \\ 
   \               &     \                   & $-125\pm 2$  & $16\pm 3$ & 13.17$\pm$0.07 \\
   \               &     \                   & $-99\pm 9$  & $10^{+19}_{-7}$ & 12.45$\pm$0.35 
\enddata

\tablenotetext{a}{For N~I, only a single, relatively weak transition
  is free of blending in the STIS spectrum, and consequently the N~I
  measurements should be treated with caution (at Outer Arm
  velocities, the N~I $\lambda \lambda$ 1200.22, 1200.71 transitions
  are blended with lower-velocity absorption in the other lines of the
  N~I triplet).  In addition, the third component that is evident in
  the other metal profiles (at $v \approx -117$ km s$^{-1}$) is too
  weak to be fitted in the N~I $\lambda$1199.55 profile.}
\tablenotetext{b}{The S~II $\lambda 1259.52$ profile is corrected for
  blending with an extragalactic O~VI absorber (see text).  The S~II
  measurements should also be treated with caution because they are
  weak and marginally detected.  Only a single component could be
  fitted to these weak lines.}
\tablenotetext{c}{As can be seen from Figure~\ref{fig_multicomp},
  there is a discrepancy between the Si~{\sc iv} $\lambda$1393.76 and
  $\lambda$1402.78 profiles at $-115 \lesssim v_{\rm LSR} \lesssim
  -80$ km s$^{-1}$.  This discrepancy is due to an {\sc H i}
  Ly$\alpha$ line at $z_{\rm abs}$ = 0.14760 that blends with the
  Galactic Si~{\sc iv} $\lambda$1393.76 profile.  The identification
  of this blend as this {\sc H i} Ly$\alpha$ line is established by
  the presence of an {\sc H i} Ly$\beta$ line at this $z$ in the {\it
    FUSE} spectrum of H1821+643 (K. R. Sembach et al., in
  preparation).}
\end{deluxetable*}

All of our targets, both stellar and quasistellar, have been observed
with the E140M or E140H echelle modes of STIS.  The QSOs have also
been observed with {\it FUSE}.  Information about the observations of
H1821+643 and HS0624+6907, and about the reduction of the data, can be
found in Tripp et al. (2001,2008) and Aracil et al. (2006).
Table~\ref{star_obs_log} provides a log of the STIS observations of
stars that we use to constrain the distance to the OA and HVC Complex
G in \S \ref{distance_section}.  The STIS spectra of these stars have
also been reduced as described in Tripp et al. (2001).  Information on
the design and performance of STIS can be found in Woodgate et
al. (1998) and Kimble et al. (1998); the design and performance of
{\it FUSE} has been discussed by Moos et al. (2000,2002) and Sahnow et
al. (2000).

The STIS observations of H1821+643 and HS0624+6907 were obtained with
the E140M echelle mode with the $0.\arcsec 2 \times 0. \arcsec 06 $
slit; this mode provides 7 km s$^{-1}$ resolution (FWHM) and covers
the $1150-1730$ \AA\ range. The FUSE data for these sight lines were
obtained with the large ($30 \arcsec \times 30 \arcsec $) aperture in
all four {\it FUSE} channels (see Moos et al. 2000), which provide a
resolution of $\approx 20$ km s$^{-1}$ and cover the $905-1187$
\AA\ range. However, the S/N is very low in the {\it FUSE} SiC spectra
of HS0624+6907, which cover $\lambda \lesssim$ 1000 \AA, so for that
sight line we only use {\it FUSE} data from the LiF channels.  To
maximize the S/N of the HS0624+6907 spectra, we combine the {\it FUSE}
spectra recorded during both the day and night sides of the {\it FUSE}
orbit.  This leads to inclusion of strong emission lines from the
Earth's atmosphere that are excited by sunlight and thus are
predominantly present on the day side of the orbit.  Fortunately, most
of the Outer Arm absorption lines of interest are at a velocity that
shifts the lines well away from the terrestrial dayglow emissions.
For H1821+643, the {\it FUSE} data have substantially higher S/N, so
we use the data from all four {\it FUSE} channels but only include the
orbital-night photons in order to suppress the strong dayglow emission
lines.  Figure~\ref{h1821vel} shows the final coadded and
continuum-normalized data for Galactic absorption lines of interest
toward H1821+643, and Figure~\ref{hs0624vel} shows Galactic absorption
lines in the spectrum of HS0624+6907.

The measurements of \ion{H}{1} 21 cm emission along the two QSO sight
lines were obtained by Wakker et al. (2001) using the Effelsberg 100m
telescope, which has a $9 \arcmin$ beam. Both of the spectra show
multiple well-defined velocity components, and we focus on the
high-velocity 21 cm components which are associated with the Outer
Arm. According to Wakker et al. (2001), the detected Outer Arm
component in the direction of H1821+643 has $v_{\rm LSR} = -128$ km
s$^{-1} $ and $N$(\ion{H}{1}) = $(3.3 \pm 0.5) \times 10^{18}$
cm$^{-2}$, and toward HS0624+6907, the Outer Arm is detected at
$v_{\rm LSR} = -100$ km s$^{-1}$ with $N$(\ion{H}{1})= $(19.8 \pm 1.8)
\times 10^{18}$ cm$^{-2}$.  However, Wakker et al. (2001) have
compared their Effelsberg data to \ion{H}{1} measurements of the same
targets with smaller beams ($1' - 2'$, see their Table 1), and they
conclude that the larger Effelsberg beam introduces systematic
uncertainties at the level of $\approx$25\% .  For this reason, we
include an additional 25\% uncertainty in these \ion{H}{1} column
densities.

\section{Absorption-Line Measurements}
\label{absmeas}

We use two techniques to extract information from the ultraviolet
absorption lines presented in this paper. First, we construct apparent
column density profiles using the apparent optical depth (AOD) method
(Savage $\&$ Sembach 1991). In brief, in this method the optical depth
of absorption in a pixel at velocity $v$ is first determined from the
usual relation, $\tau _{\rm a} (v) = {\rm ln} [I_{\rm c}(v)/I_{\rm
    ob}(v)] $, where $I_{\rm ob}(v)$ is the observed flux and $I_{\rm
  c}(v)$ is the continuum flux.  We estimate the continuum flux [and
  normalize $I_{\rm ob}(v)$] by fitting a low-order polynomial to the
adjacent continuum near an absorption line of interest; typically we
use the region within rougly $\pm 1000$ km s$^{-1}$ of the line for
continuum fitting.  The apparent column density at $v$ is then
determined from the apparent optical depth, $N_{\rm a}(v)=(m_{e}c/ \pi
e^{2})(f \lambda)^{-1} \tau _{\rm a}(v)=3.768 \times 10^{14}(f
\lambda)^{-1} \tau _{\rm a}(v)$, where $f$ is the oscillator strength
and $\lambda $ is the transition wavelength in \AA. The other symbols
have their common meanings. One virtue of this approach is that the
effects of line saturtion can be quickly and easily recognized by
comparing the $N_{\rm a}(v)$ profiles of two or more resonance lines
of a given species which differ in the product $f \lambda$.  Detailed
discussions of the use and benefits of the AOD technique can be found
in Savage \& Sembach (1991) and Jenkins (1996).

\begin{figure*}
\epsscale{1.0}
\plotone{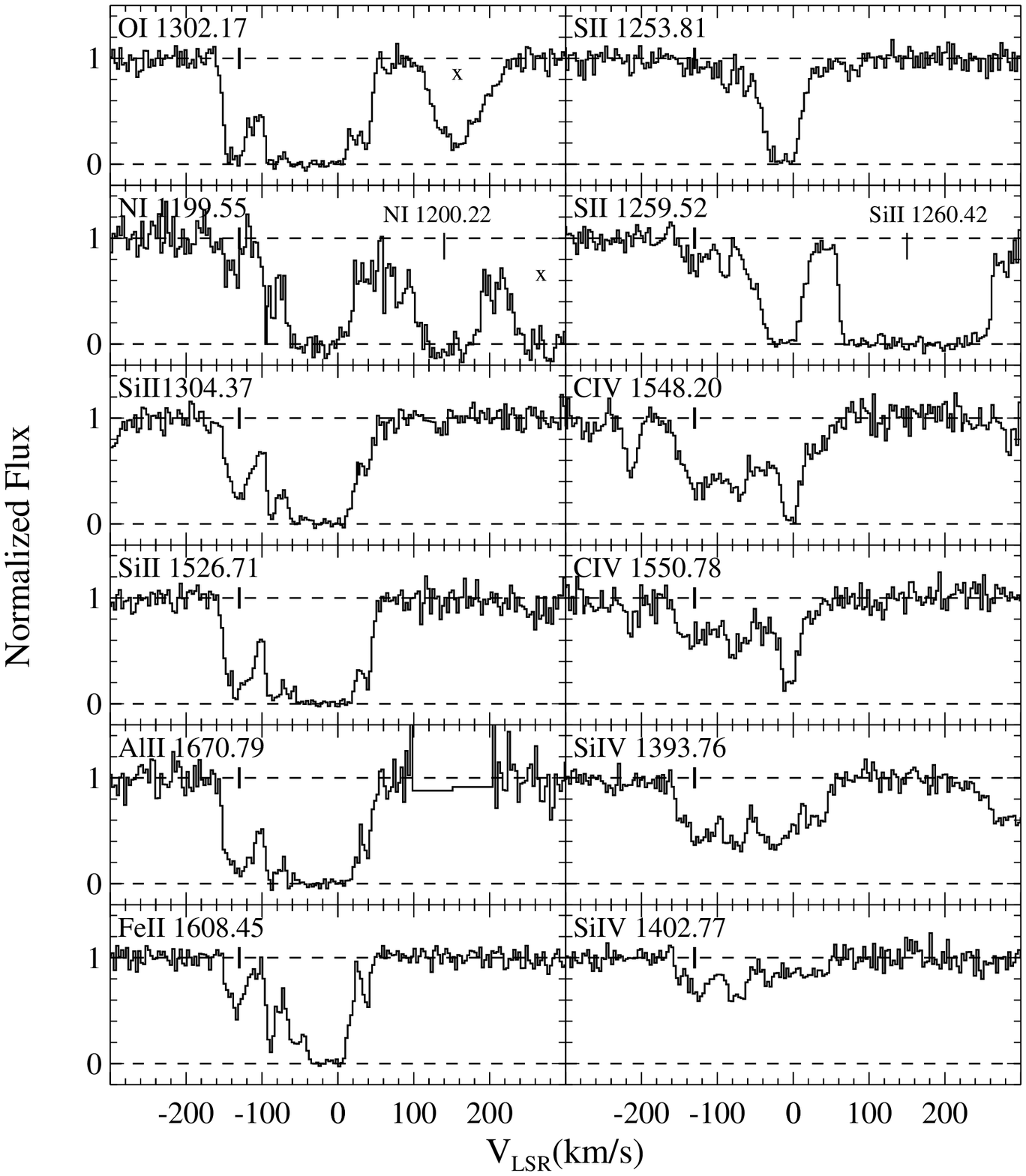}
\caption{Continuum-normalized UV absorption profiles of Galactic lines
  detected in the STIS E140M echelle spectrum of H1821+643, plotted
  vs. LSR velocity. The profile transition is indicated in each panel.
  In some panels, other Milky Way lines that happen to fall within the
  plotted velocity range are labeled; some other unrelated absorption
  features are marked with an 'x'. The Al~{\sc ii} $\lambda$1670.79
  line is near a gap between the STIS echelle orders, and there is no
  data between $v = 100$ and 200 km s$^{-1}$.  The Outer Arm
  components, at $v_{\rm LSR} \approx -130$ km s$^{-1}$, are marked
  with a vertical tick mark. \label{h1821vel}}
\end{figure*}

\begin{deluxetable*}{lcccc}
\tablewidth{0pc}
\tablecaption{Outer-Arm Profile-Fitting Measurements: HS0624+6907\label{tab:hs0624}}
\tablehead{Species \ \ \ & Fitted Lines & $v$ (LSR) & b & log [$N$ (cm$^{-2}$)] \\
             \     & (\AA )       & (km s$^{-1}$) & (km s$^{-1}$) & \ }
\startdata
N~I\dotfill & 1199.55,1200.22\tablenotemark{a} & $-104:$ & $5:$ & $\gtrsim 14.6:$ \\
Si~II\dotfill & 1190.42,1193.29,1260.42 & $-101:$ & $11:$ & $\gtrsim 14.4:$  \\
   & 1304.37,1526.71\tablenotemark{a} & & & \\
S~II\dotfill & 1253.81,1259.52 & $-100 \pm 1$&$7 \pm 1$ &$14.18 \pm 0.07$ \\
Al~II\dotfill & 1670.79\tablenotemark{a} & $-104:$ & $10:$ & $\gtrsim 13.00:$  \\
Fe~II\dotfill & 1121.98,1143.23,1144.94& $-104 \pm 1$ & $8 \pm 1$& $14.03 \pm 0.06$ \\
   &  1608.45 & & &  \\
C~IV\dotfill & 1548.20,1550.78 & $-107 \pm 1$ & $5 \pm 1$& $13.07 \pm 0.11$ \\
Si~IV\dotfill & $1393.76,1402.77$ & $-101 \pm 1$ & $16 \pm 1$& $13.06 \pm 0.03$ \\
\enddata
\tablenotetext{a}{Due to line saturation, these measurements are
  highly uncertain.}
\end{deluxetable*}

\begin{figure*}
\epsscale{1.0}
\plotone{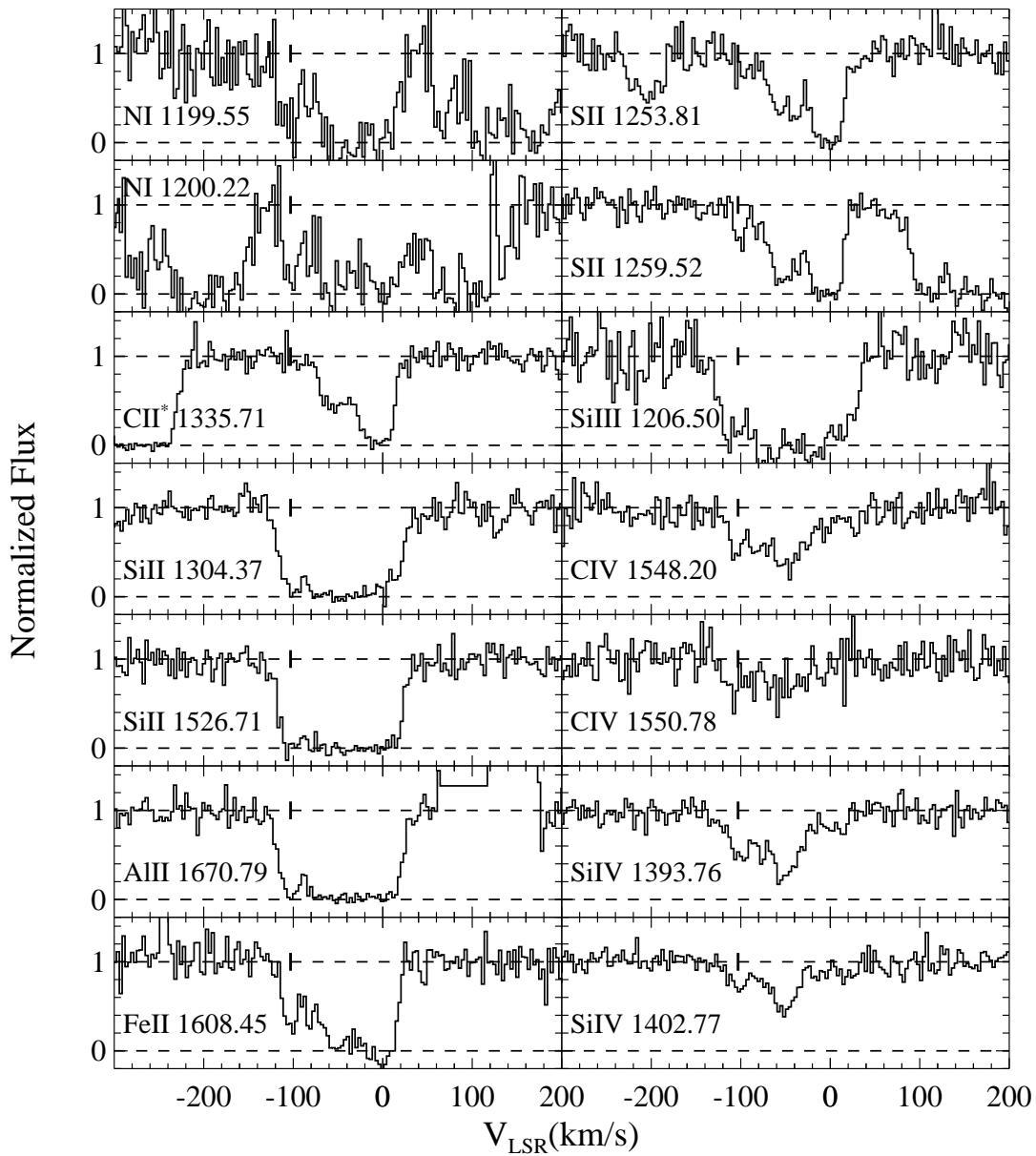}
\caption{UV absorption lines detected in the STIS E140M echelle
  spectrum of HS0624+6907, as in Figure~\ref{h1821vel}. The high
  velocity components affiliated with the Outer Arm are indicated with
  vertical tick marks, at $v_{\rm LSR} \approx -100$ km
  s$^{-1}$.  \label{hs0624vel}}
\end{figure*}

We will find the $N_{\rm a}(v)$ method to be illustrative, but as we
can see from Figures~\ref{h1821vel} - \ref{hs0624vel}, many of the
Outer Arm absorption lines of interest are blended with adjacent
components at lower velocities, so we need to be able to deblend these
features.  Moreover, we will show evidence below that even the OA
absorption itself is composed of closely spaced narrow components.
Consequently, we use the Voigt-profile fitting software of Fitzpatrick
\& Spitzer (1997), including the effects of the STIS line-spread
function (Proffitt et al. 2010), to measure the velocity centroids,
column densities, and line widths (expressed as $b-$values) of the
lines.
 
\subsection{H1821+643}
\label{h1821measurements}

\begin{figure}
\epsscale{1.20}
\plotone{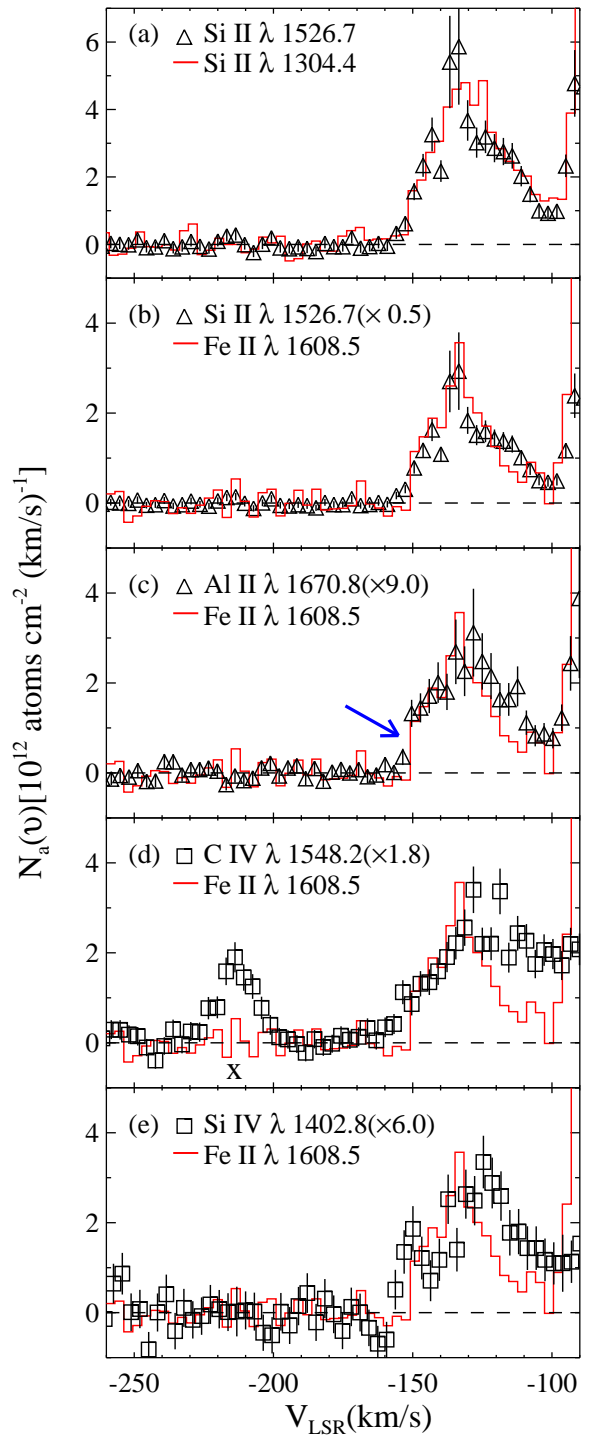}
\caption{Comparison of various apparent column density profiles (\S
  \ref{absmeas}) observed toward H1821+643 in the Outer Arm, plotted
  vs. LSR velocity. The panels show (a) Si~{\sc ii} $\lambda 1304.4$
  (red histogram) vs. Si~{\sc ii} $\lambda 1526.7$ (triangles with
  $1\sigma$ error bars)), (b) Fe~{\sc ii} $\lambda$1608.5 (red
  histogram) vs. Si~{\sc ii} $\lambda 1526.7$ ($\times$0.5,
  triangles), (c) Fe~{\sc ii} $\lambda$1608.5 (red histogram)
  vs. Al~{\sc ii} $\lambda$1670.8 ($\times$9.0, triangles), (d)
  Fe~{\sc ii} $\lambda$1608.5 (red histogram) vs. {\sc C iv}
  $\lambda$1548.2 ($\times$1.8, squares), and (e) Fe~{\sc ii}
  $\lambda$1608.5 (red histogram) vs. Si~{\sc iv} $\lambda$1402.8
  ($\times$6.0, squares).  A sharp edge is consistently present on the
  blue side of many of the profiles (indicated with a blue arrow in
  the Fe~{\sc ii} vs. Al~{\sc ii} comparison). The feature at $v_{\rm
    LSR} = -213$ km s$^{-1}$ is due to \ion{C}{4} and is confirmed by
  the other line of the \ion{C}{4} doublet, but no other metals are
  detected at that velocity, with the possible exception of \ion{O}{6}
  (see Tripp et al. 2003).\label{h1821nav}}
\end{figure}

\begin{figure}
\epsscale{1.3}
\plotone{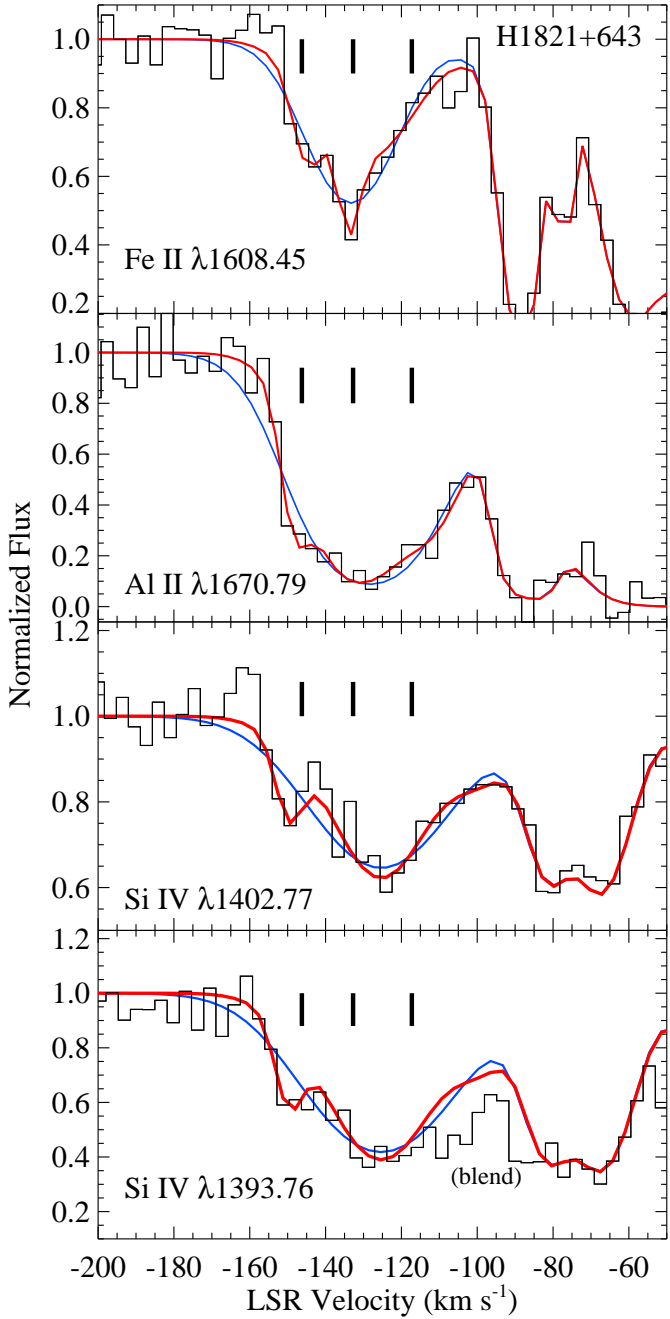}
\caption{Comparison of a Voigt-profile fit to the H1821+643 Outer Arm
  component (at $v \approx -130$ km s$^{-1}$) assuming a single
  component for the OA (smooth blue line) or 3 components for the OA
  (smooth red line).  The fits are overplotted on the observed
  profiles (histograms) of Fe~{\sc ii} $\lambda 1608.45$, Al~{\sc ii}
  $\lambda 1670.79$, and Si~{\sc iv} $\lambda \lambda 1393.76,
  1402.77$, as labeled in each panel.  For components at velocities
  outside of the OA velocity range, the two models have identical
  component structure. Comparison of the one-component and
  three-component fits show that the three-component fit provides a
  better fit to the detailed sharp features that consistently appear
  in various profiles (as also shown in
  Figure~\ref{h1821nav}).\label{fig_multicomp}}
\end{figure}

The Outer Arm absorption toward H1821+643 has been studied previously
by Savage et al. (1995) and Tripp et al. (2003).  However, the initial
investigations assumed a single component for the OA absorption
profile and only employed a subset of the currently available lines
detected in the OA, so it is worthwhile to revisit these data.  

Careful examination of the STIS H1821+643 data reveals that the OA
component structure is more complicated than a single component.  To
show this, we compare the apparent column density profiles of selected
species in Figure~\ref{h1821nav}, and we plot expanded absorption
profiles in Figure~\ref{fig_multicomp}.  Several features in these
figures provide evidence of multiple components in the OA velocity
range.  First, the $N_{\rm a}(v)$ profiles consistently show a sharp
edge at $v_{\rm LSR} = -150$ km s$^{-1}$ in several different species.
For example, both of the \ion{Si}{2} profiles in panel (a) of
Figure~\ref{h1821nav} show a consistent edge at this $v$, and the
\ion{Fe}{2} and \ion{Al}{2} profiles (panel c) exhibit the same
feature.  Such a sharp discontinuity cannot occur in a profile due to
a single Voigt component.  Instead, this feature requires at least one
narrow component near $v_{\rm LSR} \approx -150$ km s$^{-1}$ that is
blended with another component at $v_{\rm LSR} > -150$ km s$^{-1}$.
In fact, we can see from the \ion{Fe}{2} $\lambda$1608.45 and
\ion{Si}{2} $\lambda$1526.71 lines (panel b) that there are
indications of three components in the OA: a narrow feature at $v_{\rm
  LSR} \approx -135$ km s$^{-1}$ and a broader component at $v_{\rm
  LSR} \approx -125$ kms$^{-1}$ (in addition to the component causing
the sharp edge at $v_{\rm LSR} \approx -150$ km s$^{-1}$).  To show
this a different way, we compare in Figure~\ref{fig_multicomp} a
single-component fit to a three-component fit of the OA absorption
profiles of \ion{Fe}{2} $\lambda$1608.45 and \ion{Al}{2}
$\lambda$1670.79.  The three-component fit is superior, both for
fitting the sharp edge at $v_{\rm LSR} \approx -150$ km s$^{-1}$ and
for fitting the detailed component structure at $v_{\rm LSR} > -150$
kms$^{-1}$.  The $\chi ^{2}$ statistics for the fits indicate that the
one-component fits are acceptable but the three-component fits are
better.  For example, for the \ion{Fe}{2} $\lambda$1608.45 line, the
reduced $\chi ^{2}$ for the single-component fit is $\chi ^{2}_{\nu}$
= 1.08 while the three-component fit has $\chi ^{2}_{\nu}$ = 0.93. We
note that we have also explored whether two-component fits might be
favored.  We find that the two-component fits provide no improvement
compared to one-component fits; in this case, the profile-fitting code
converges to a solution that is virtually identical with the
one-component model with the same $\chi ^{2}_{\nu}$.  Three components
are required to improve the fit to the sharp edge and the narrow core
of the profiles.

Based on the consistent evidence in multiple profiles, we have elected
to revise the profile fits published previously, and our results from
fitting three components to the absorption lines detected in the OA
are summarized in Table~\ref{tab:h1821}.  Interestingly, many of the
line widths indicated by the fits are relatively narrow, which
potentially has implications regarding the physical conditions of the
gas (\S \ref{ionabun}). We note that it can be difficult to extract
reliable line widths from strongly blended components, but the
presence of the sharp edge in many of the absorption profiles requires
a narrow component in many of the species.  This alone has interesting
implications about the nature of the OA.  We note that most of the OA
metals in Table~\ref{tab:h1821} are not saturated, and the column
densities summed over the three components or determined from a single
component are quite similar and are robust.

Perhaps even more interesting is the indication of a similarly narrow
component in the profiles of the highly ionized species \ion{Si}{4}
and \ion{C}{4} (see Figure~\ref{h1821vel}).  A narrow feature can be
seen consistently in the \ion{Si}{4} $\lambda \lambda$1393.76, 1402.77
doublet at $v_{\rm LSR} = -150$ km s$^{-1}$; to show this we again
plot the \ion{Si}{4} doublet, with single-component and a
three-component fits overlaid, in the lowest panels in
Figure~\ref{fig_multicomp}. In the high ions, this narrow feature is
offset to somewhat more negative velocities than in the low ions, but
it is clearly present.  This is interesting because in collisionally
ionized gas, species ionized to this degree should be much broader due
the higher temperature of the plasma.  We will return to this issue in
\S \ref{ionabun}.

We note that in the spectrum of H1821+643, the measurement of
\ion{S}{2} in the Outer Arm is somewhat complicated by blending with
an extragalactic \ion{O}{6} doublet at $z_{\rm abs}$ = 0.21331 (Tripp
et al. 2008); the \ion{O}{6} $\lambda$1037.62 line at this $z$ is
blended with the \ion{S}{2} 1259.52 transition at the Outer Arm
velocity.  The presence of the \ion{O}{6} $\lambda$1037.62 in the
blend is indicated by a comparison of the \ion{S}{2} $\lambda$1253.83
and $\lambda$1259.52 $N_{\rm a}(v)$ profiles: the 1259.52 \AA\ line
indicates a greater apparent column than the 1253.83 \AA\ transition,
which is unphysical (the lines should indicate the same column or, if
there is some saturation, the $\lambda$1253.83 line should be greater
than $\lambda$1259.52).  This discrepancy is caused by extra optical
depth in the $\lambda$1259.52 profile contributed by the extragalactic
\ion{O}{6} line.  To overcome this problem, we used the \ion{O}{6}
$\lambda$1031.93 line at $z_{\rm abs}$ = 0.21331, which is free from
blending, to predict the profile of the corresponding \ion{O}{6}
$\lambda$1037.62 line, and then we divided the predicted \ion{O}{6}
$\lambda$1037.62 profile out of the \ion{O}{6} + \ion{S}{2} blend.
After removing the extragalactic \ion{O}{6}, we found the $N_{\rm
  a}(v)$ profiles of \ion{S}{2} $\lambda$1253.83 and $\lambda$1259.52
to be in good agreement.

\subsection{HS0624+6907}
\label{hs0624meas}

\begin{deluxetable*}{lcccccccc}
\tabletypesize{\scriptsize}
\tablecaption{Heliocentric and Galactocentric Distances of Stars in the Direction of the Outer Arm and Adjacent HVCs\label{distance_tab}}
\tablehead{Star & \multicolumn{2}{c}{\underline{ \ \ Galactic Coordinates \ \ }} & Distance\tablenotemark{a} & Galactocentric Radius\tablenotemark{b} & z & Reference\tablenotemark{c} & log $N$({\sc C ii}) & log $N$(Si~{\sc ii}) \\
 \ & Long. & Lat. & (kpc)    & (kpc)                 & (kpc) & \ & \ & }
\startdata
\multicolumn{9}{c}{\underline{\ \ \ \ Outer Arm Nondetections \ \ \ \ }} \\
K1$-$16\dotfill   & 94.03  & +27.43 & 1.6  & 8.7  & 0.7  & 1 &\nodata &\nodata \\
HD198781\dotfill  & 99.94  & +12.61 & 0.69 & 8.6  & 0.2  & 2 & $<$12.31\tablenotemark{d} & $<$12.58\tablenotemark{d} \\
HD203374\dotfill  & 100.51 & +8.62  & 0.34 & 8.6  & 0.1  & 3 & $\leq$12.88\tablenotemark{e} & $<$12.41\tablenotemark{d} \\ 
HD207198\dotfill  & 103.14 & +6.99  & 1.3  & 8.9  & 0.2  & 2 & $<$12.23\tablenotemark{d} & $<$12.52\tablenotemark{d} \\
HD208440\dotfill  & 104.03 & +6.44  & 1.1  & 8.8  & 0.1  & 4 & $<$12.42\tablenotemark{d} & $<$12.64\tablenotemark{d} \\
HD209339\dotfill  & 104.58 & +5.87  & 1.2  & 8.9  & 0.1  & 4 & $<$12.40\tablenotemark{d} & $<$12.33\tablenotemark{d} \\
HD201908\dotfill  & 112.40 & +20.19 & 0.13 & 8.6  & 0.04 & 5 & $<$12.34\tablenotemark{d} & $<$12.52\tablenotemark{d} \\
HD40893\dotfill   & 180.09 & +4.34  & 3.1  & 11.6 & 0.2  & 4 & $<$12.35\tablenotemark{d} & $<$12.36\tablenotemark{d} \\
\hline
\multicolumn{9}{c}{\underline{\ \ Outer Arm Detections  \ }}\\
HS1914+7139\dotfill & 102.99 & +23.91 & 14.9 & 17.6 & 6.04 & 6,7 & \nodata & $\approx$13.6 \\ 
HD43818\dotfill   & 188.49 & +3.87  & 1.9  & 10.4 & 0.1  & 4 & 13.50$\pm$0.02 &12.84$\pm$0.11  \\
\multicolumn{9}{c}{\underline{\ \ \ \ Complex G Detections \ \ \ \ }} \\
BD+35 4258\dotfill & 77.19 & $-4.74$ & 2.9 & 8.4  & $-0.2$ & 8 & 14.04$\pm$0.02 & 12.43$\pm$0.14 \\
HD210809\dotfill  & 99.85  & $-3.13$ & 4.3 & 10.2  & $-0.2$ & 2 & 13.70$\pm$0.01 & 11.82$\pm$0.03 \\
\enddata
\tablenotetext{a}{Heliocentric distance to the star.}

\tablenotetext{b}{Galactocentric radius, $R_{\rm G}^{2} = R_{0}^{2} + (r \ 
  {\rm cos} \ b)^{2} - 2rR_{0} {\rm cos} \ l \ {\rm cos} \ b$, where $(l,b)$
  are the Galactic coordinates and $r$ is the Heliocentric distance of
  the target, assuming the distance from the Sun to the Galactic
  center $R_{0}$ = 8.5 kpc. }

\tablenotetext{c}{Source of the Heliocentric stellar distance: (1)
  Oegerle et al. (2000), (2) Bowen et al. (2008), (3) Jenkins \& Tripp
  (2001), (4) Jenkins \& Tripp (2011), (5) Fitzpatrick \& Massa
  (2005), (6) Lehner \& Howk (2010), (7) Ramspeck et al. (2001), (8)
  Jenkins (2009). As discussed in Appendix B of Bowen et al. (2008),
  the uncertainties in the stellar distances range from 10\% to 30\%
  .}

\tablenotetext{d}{Upper limit derived from the $3\sigma$ equivalent
  width limit, assuming the line is on the linear part of the curve of
  growth.}

\tablenotetext{e}{A marginally significant feature is present in the
  Outer Arm velocity range toward HD203374.  However, the feature is
  very broad and shallow and is highly sensitive to continuum
  placement -- a slightly lower continuum placement would mostly
  remove the feature.  In addition, the feature is not corroborated by
  other strong interstellar lines, e.g., the Si~III profile shows no
  evidence of absorption in this velocity range.  Given the lack of
  corroborating evidence and marginal significance of the line, we
  treat this feature as an upper limit.}

\end{deluxetable*}

\begin{figure*}
    \includegraphics[width=9.0cm, angle=0]{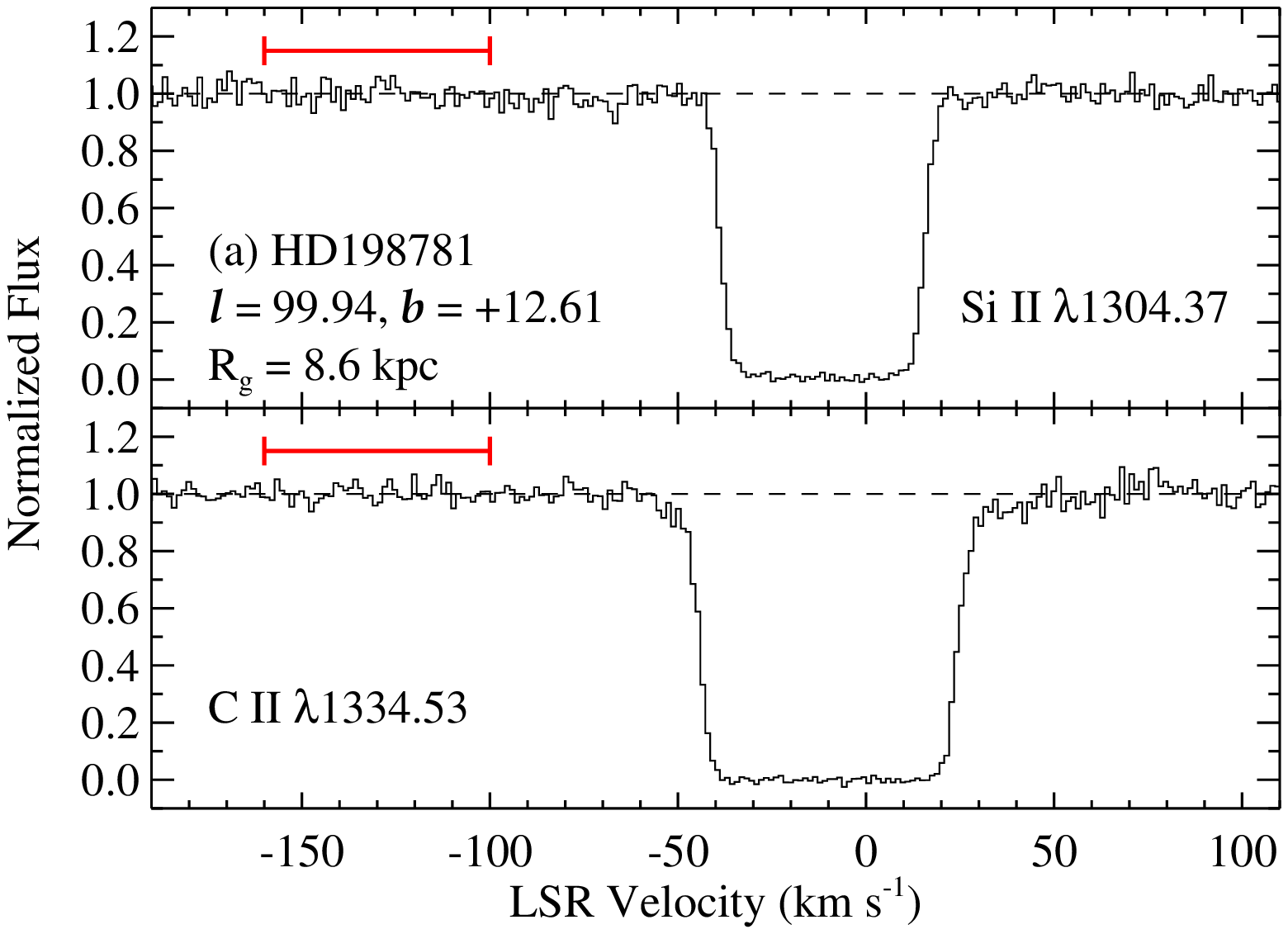}
    \includegraphics[width=9.0cm, angle=0]{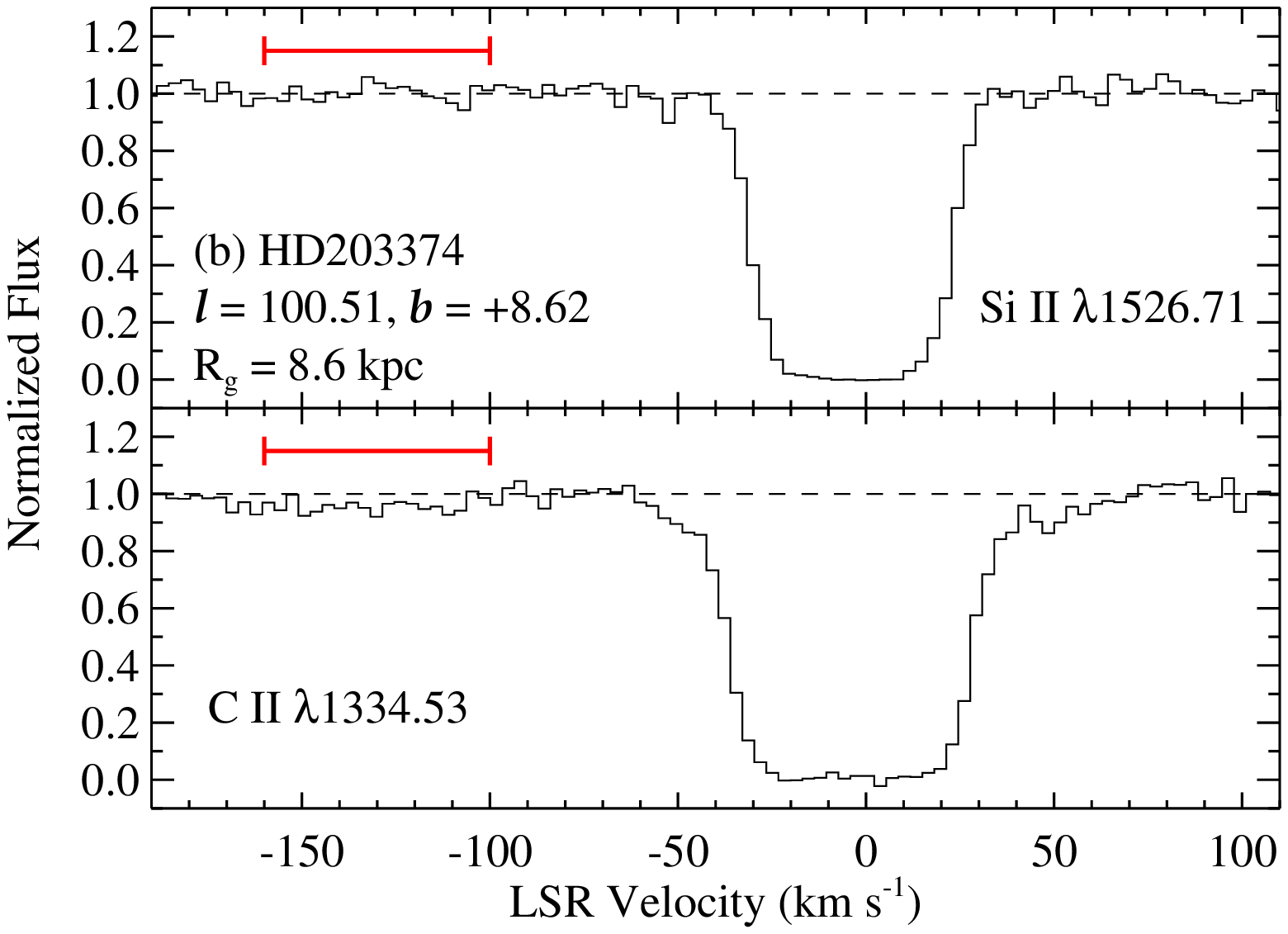}
    \includegraphics[width=9.0cm, angle=0]{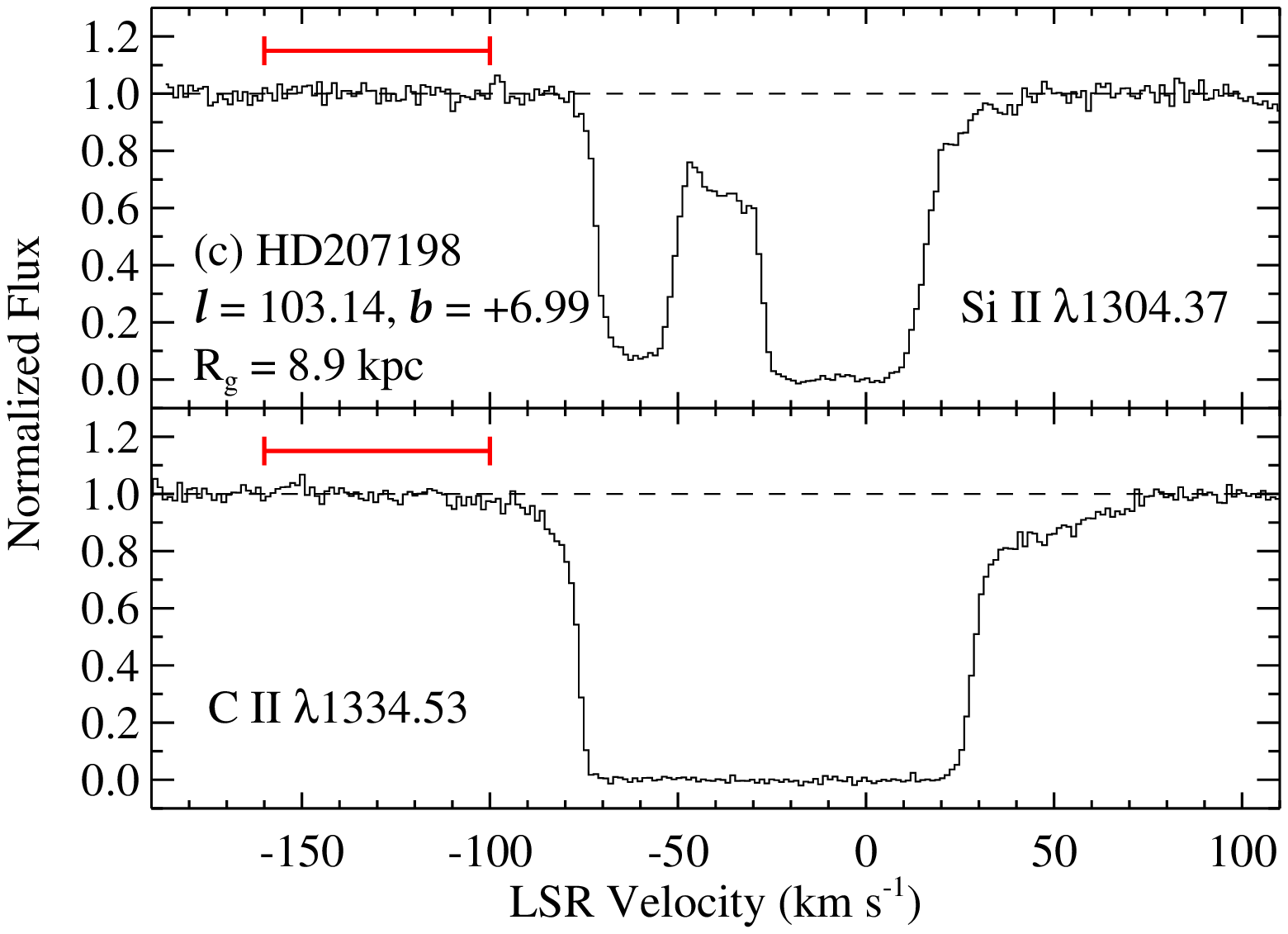}
    \includegraphics[width=9.0cm, angle=0]{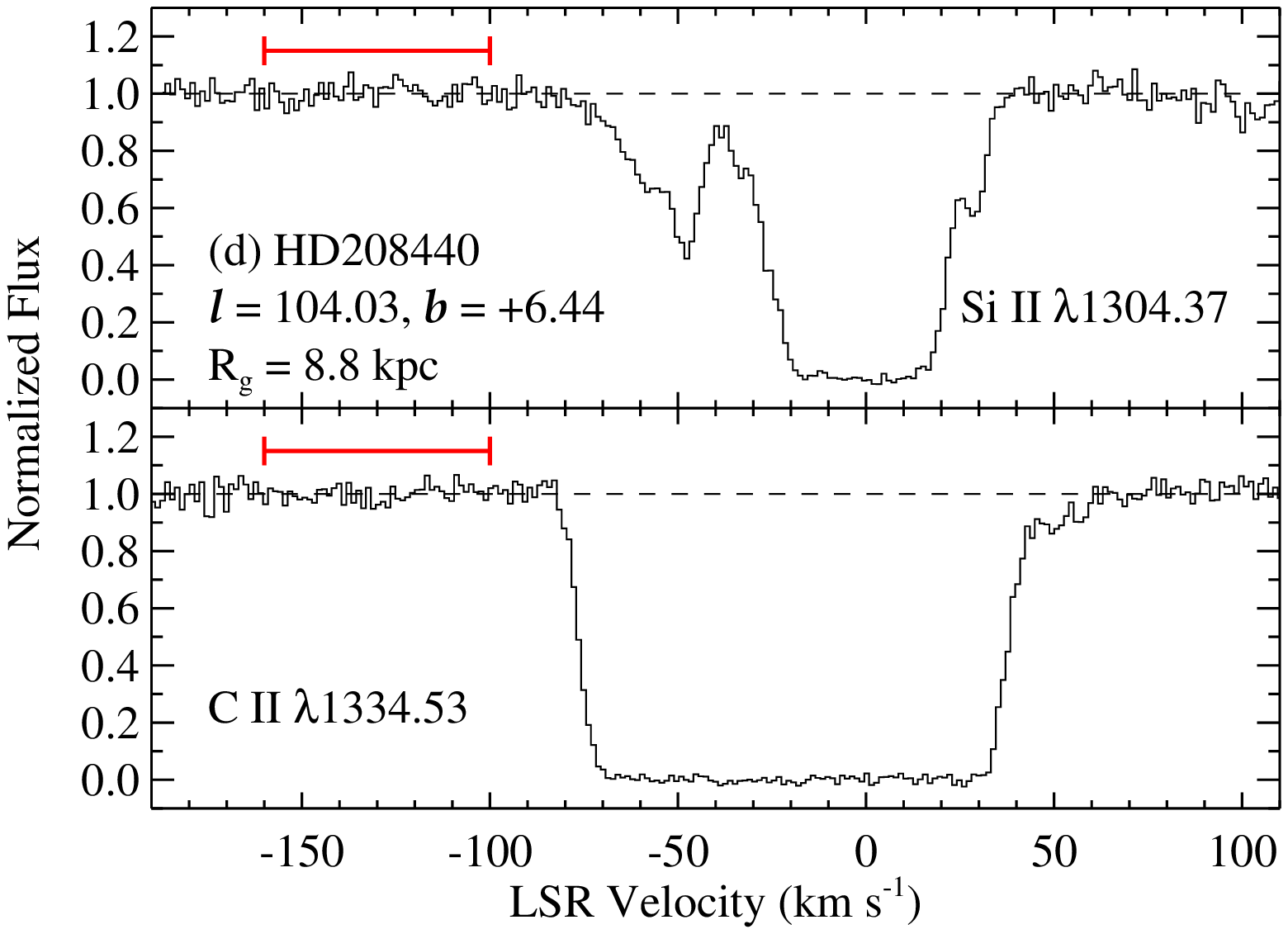}
\caption{Continuum-normalized STIS absorption profiles of strong
  interstellar lines in the spectra of the stars in the direction of
  the Outer Arm from Table~\ref{distance_tab}: (a) HD198781, (b)
  HD203374, (c) HD207198, (d) HD208440, (e) HD209339, (f) HD201908,
  (g) HD40893, and (h) 43818.  For comparison, absorption profiles of
  the same species from the spectra of H1821+643 and HS0624+6907 are
  shown in (i) and (j), respectively. For each sight line, the upper
  panel shows one of the Si~\textsc{ii} absorption profiles and the
  lower panel shows the C~\textsc{ii} $\lambda$1334.53 line.  The
  horizontal red bar in each upper panel indicates the velocity range
  over which OA absorption is detected toward H1821+643, except in the
  panels of HD40893 and HD43818; these stars are closer to
  HS0624+6907, so in these panels (g and h) the red bar shows the
  velocity range of the OA toward HS0624+6907.  Despite high S/N
  ratios, only one star, HD43818, shows UV absorption in the velocity
  range of the OA (see Figure~\ref{hd43818detect}).  Several
  intermediate-velocity absorption components detected toward
  H1821+643 are also evident toward HD207198, HD208440, and HD209339;
  these are likely to be affiliated with the Perseus
  arm. \label{starlimits}}
\end{figure*}

\begin{figure*}
\figurenum{9}
    \includegraphics[width=9.0cm, angle=0]{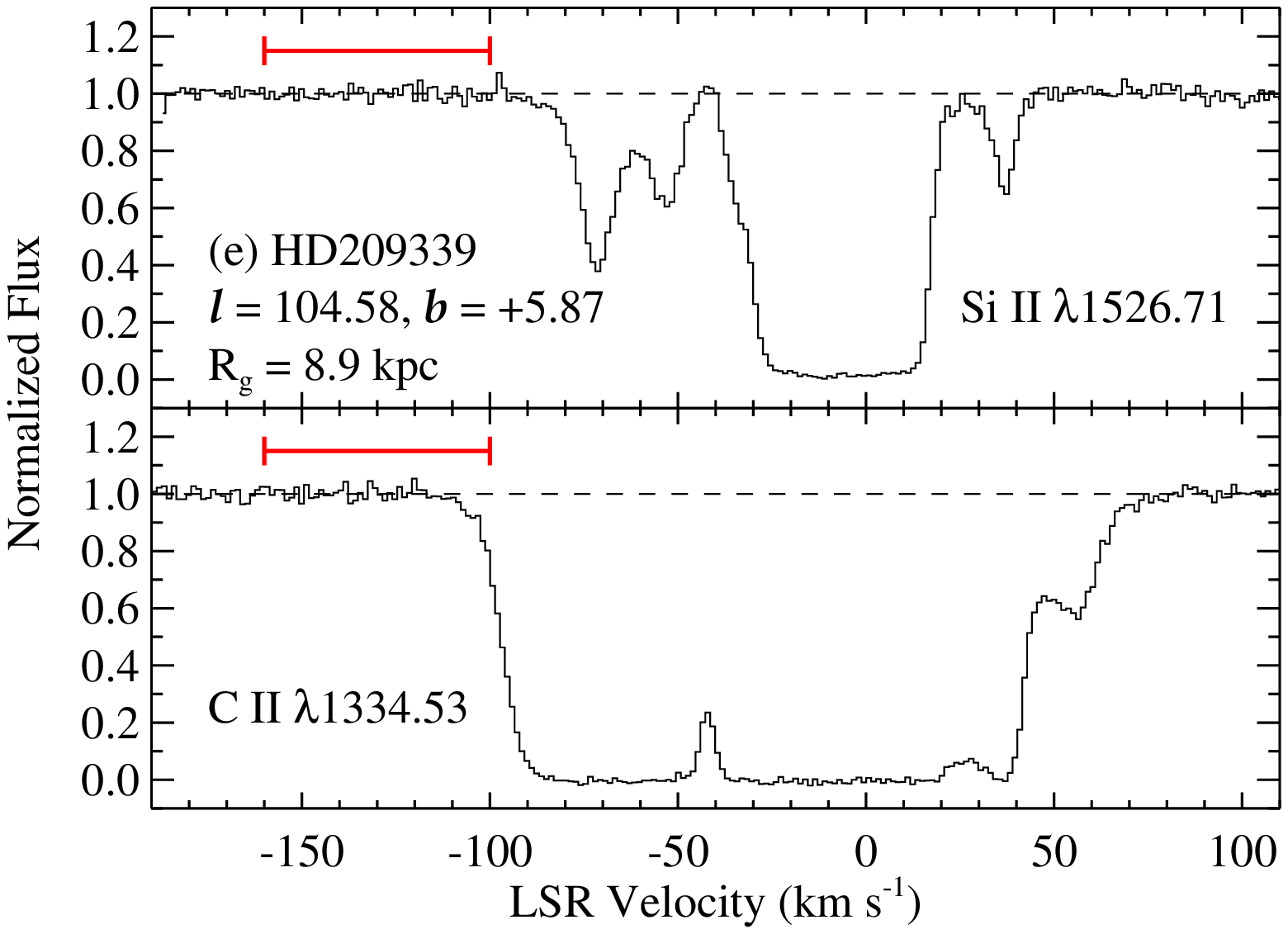}
    \includegraphics[width=9.0cm, angle=0]{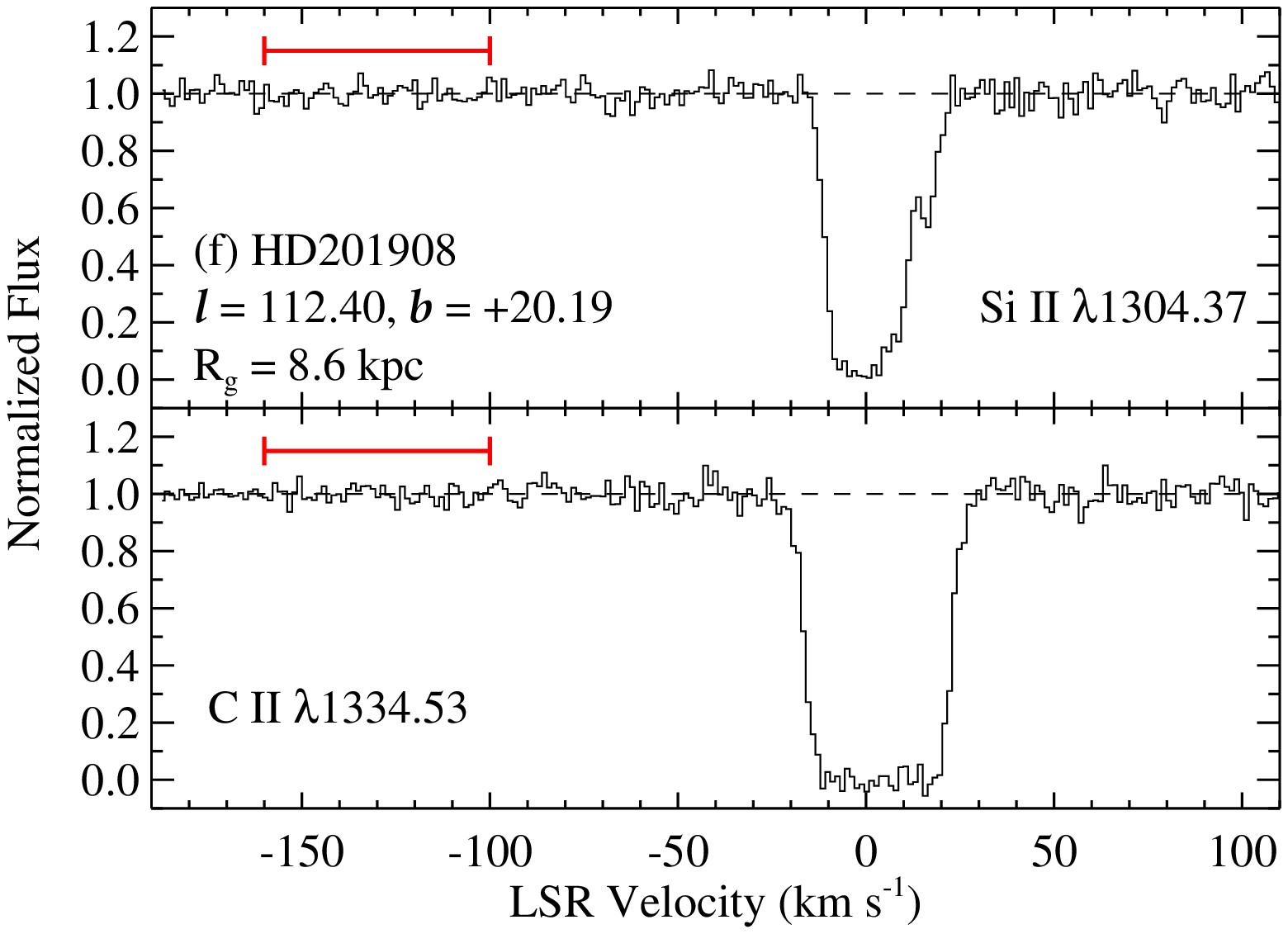}
    \includegraphics[width=9.0cm, angle=0]{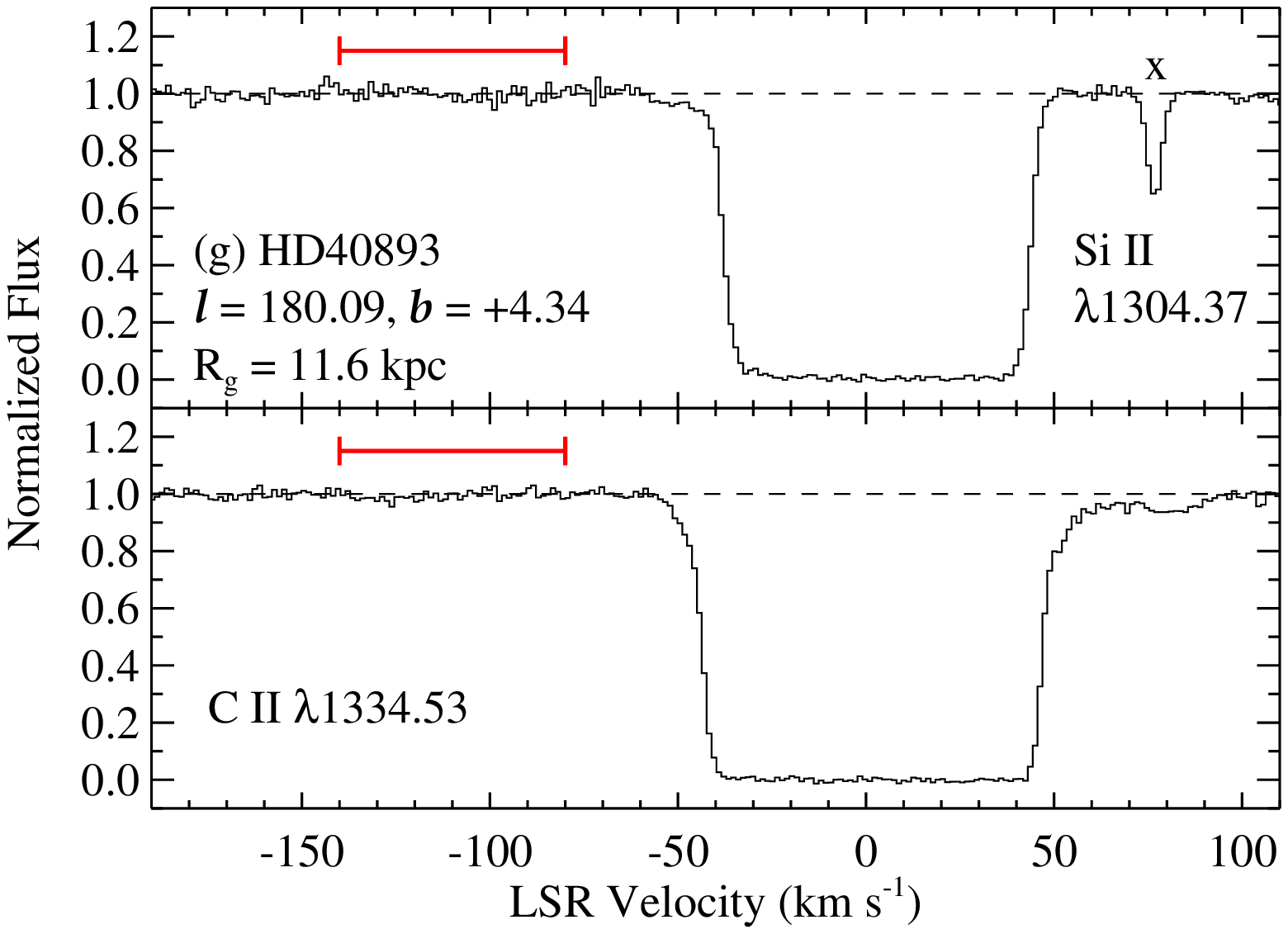}
    \includegraphics[width=9.0cm, angle=0]{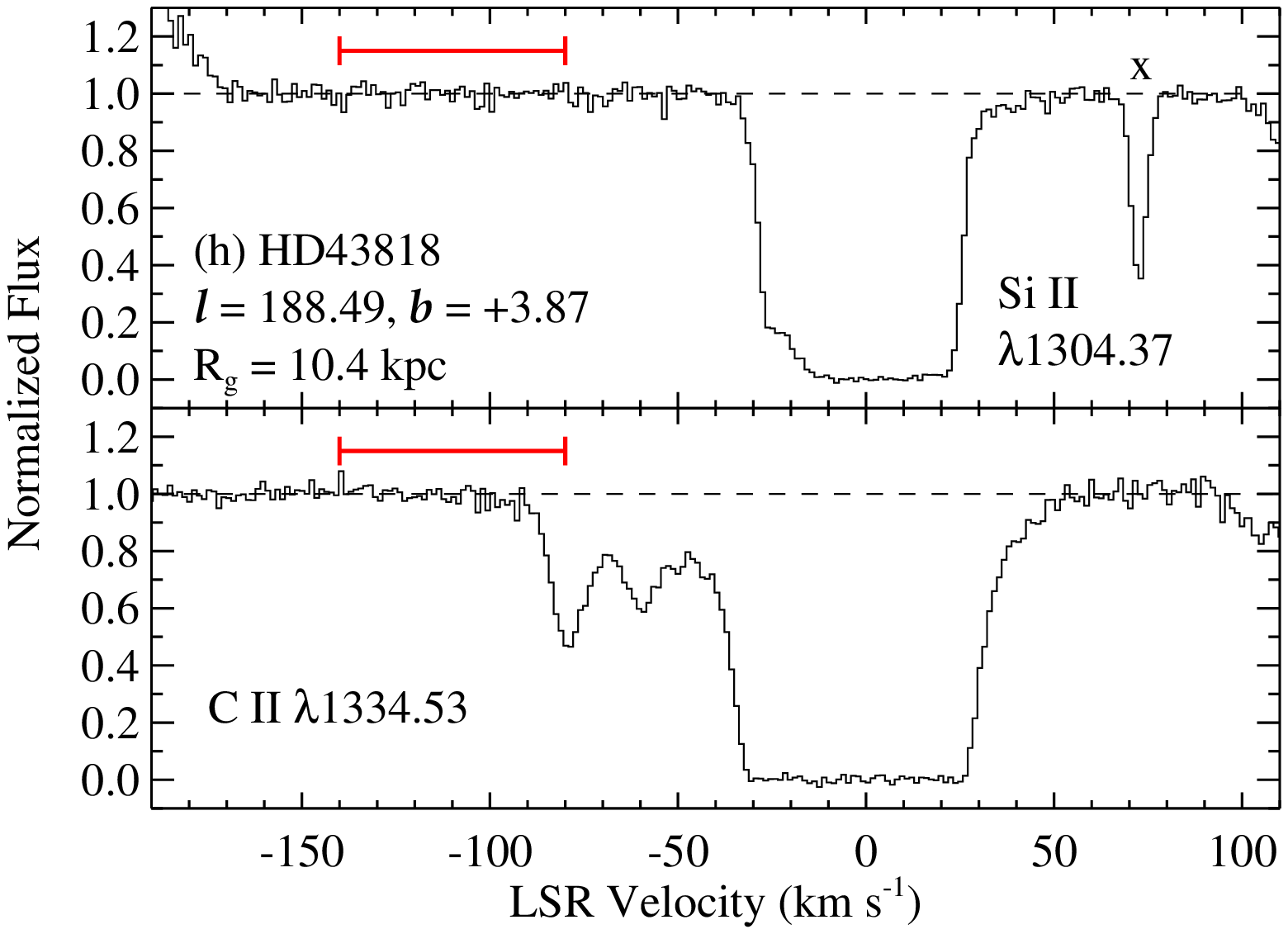}
    \includegraphics[width=9.0cm, angle=0]{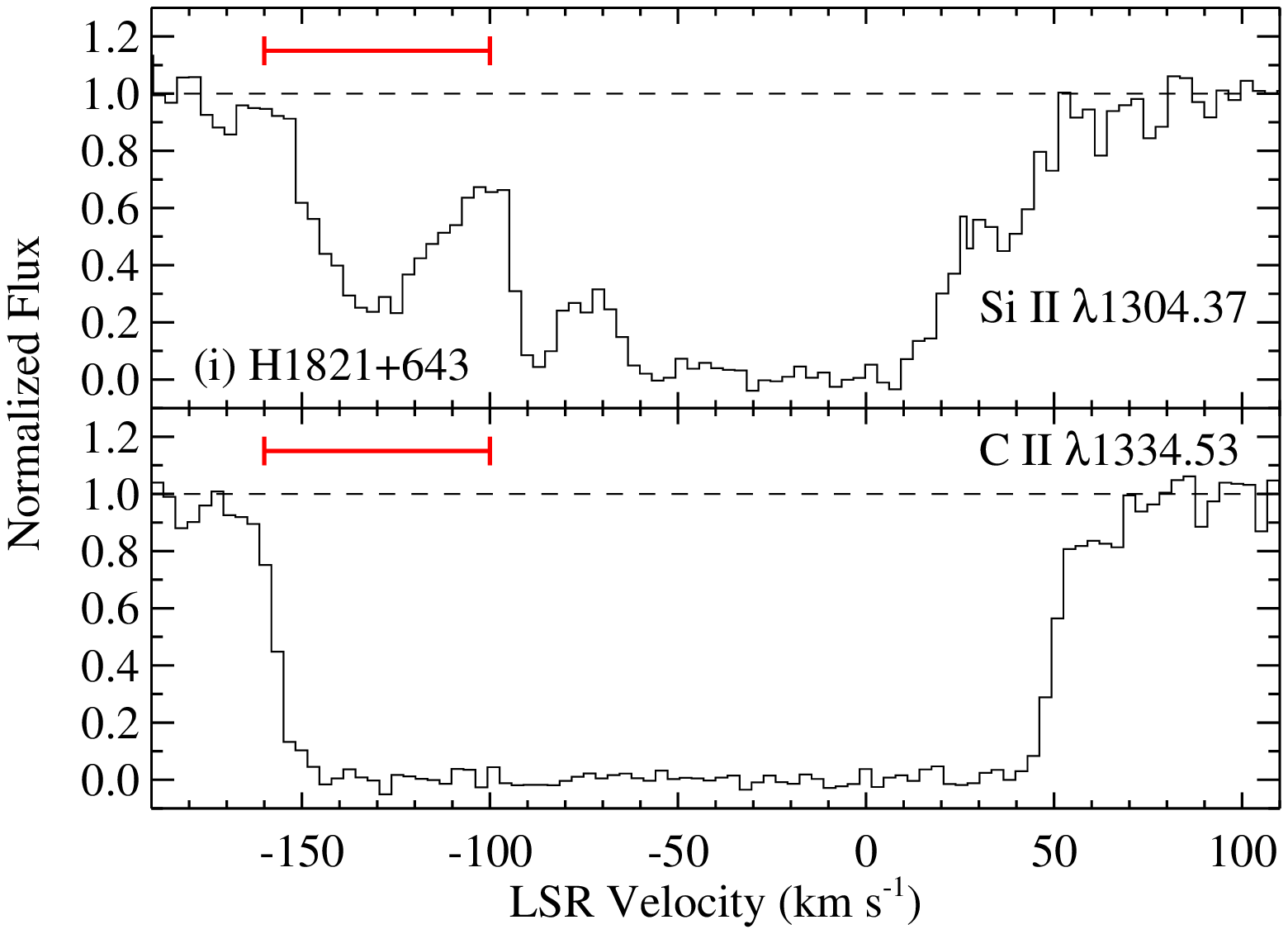}
    \includegraphics[width=9.0cm, angle=0]{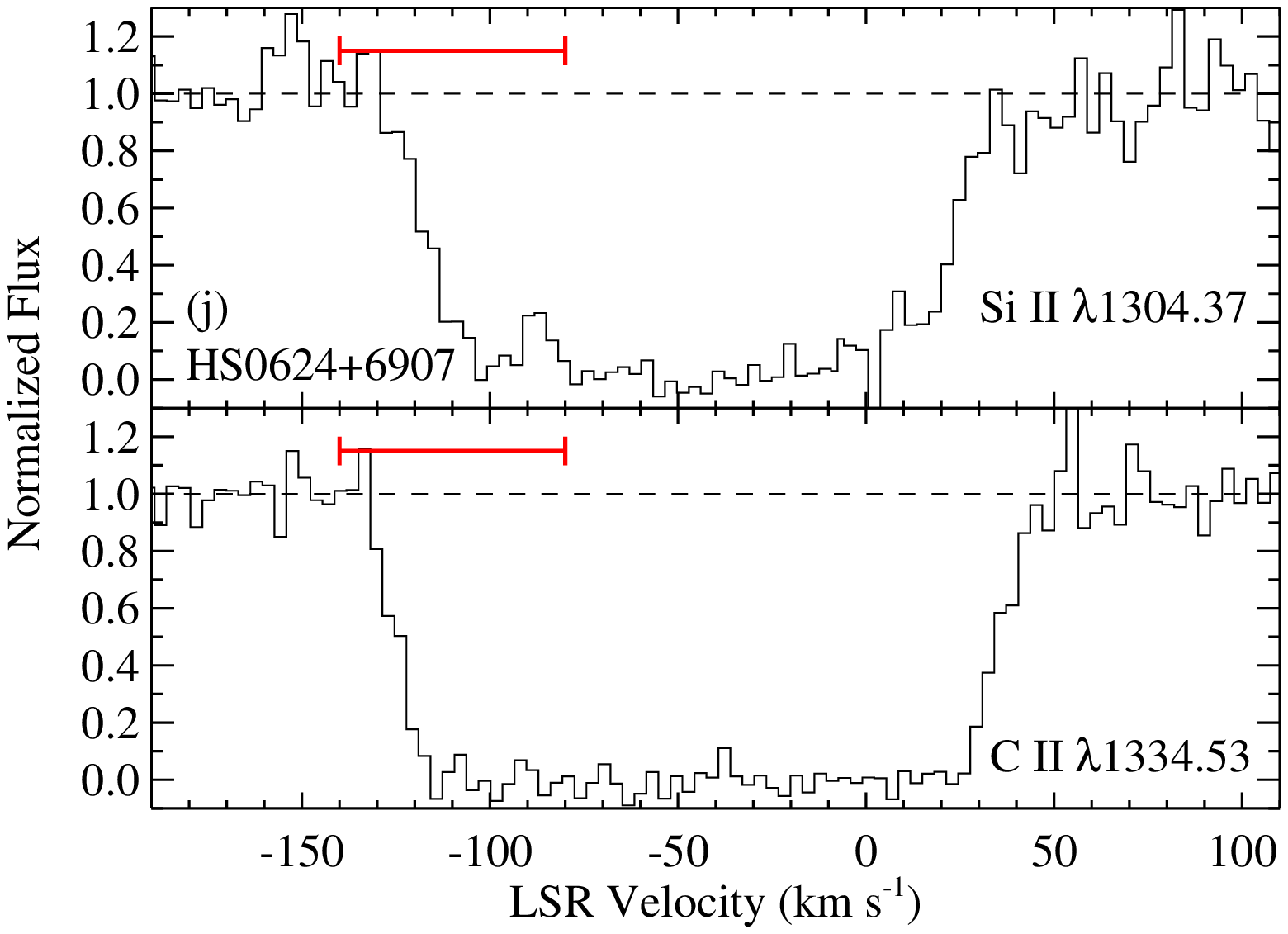}
\caption{(continued)}
\end{figure*}

\begin{figure}
\epsscale{1.3}
\plotone{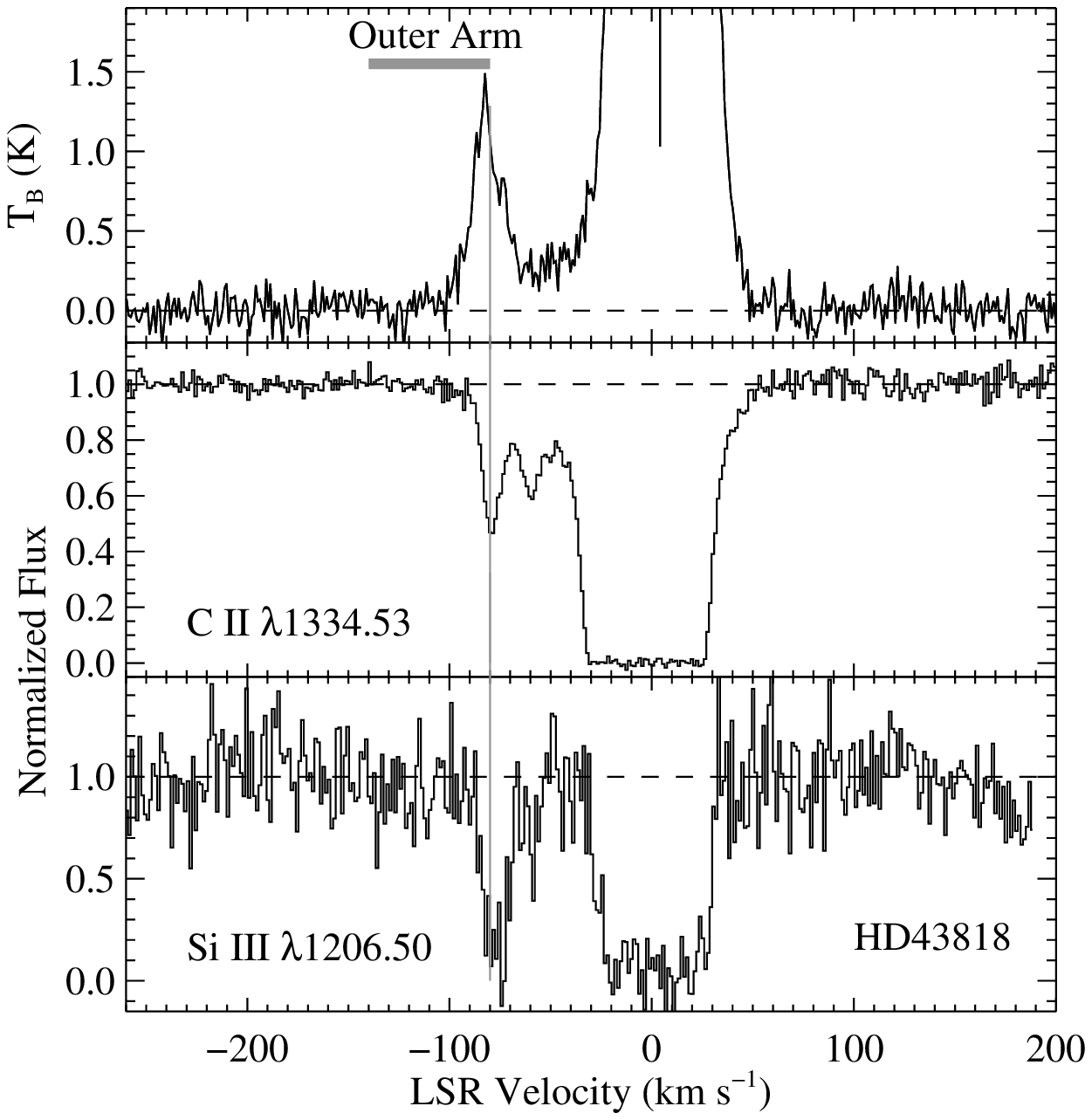}
\caption{Comparison of the 21cm emission (from the LAB survey) in the
  direction of HD43818 (upper panel) to the ultraviolet absorption
  profiles of C~\textsc{ii} $\lambda 1334.53$ and Si~\textsc{iii}
  $\lambda 1206.50$ recorded with STIS (middle and lower panels,
  respectively). Significant C~\textsc{ii} and Si~\textsc{iii}
  absorption lines are clearly detected at the velocity of the OA 21
  cm emission.\label{hd43818detect}}
\end{figure}

Toward HS0624+6907, the ultraviolet absorption in the Outer Arm
concentrates near $v_{\rm LSR} = -100$ km s$^{-1}$ (see
Figure~\ref{hs0624vel}), in agreement with the 21 cm emission velocity
reported by Wakker et al. (2001).  The measurements of the OA
absorption lines in the spectrum of this QSO are summarized in
Table~\ref{tab:hs0624}.  Unfortunately, because the gas column density
is substantially higher toward HS0624+6907, many of the species that
imprint absorption on the HS0624+6907 spectrum at OA velocities are
difficult to measure due to line saturation.  The {\it FUSE} SiC data
do not have adequate S/N, so the only transitions available for
\ion{O}{1} are strongly saturated and do not yield useful
measurements.  The \ion{N}{1} triplet at $\approx$ 1200 \AA\ is
weaker, but comparison of the \ion{N}{1} $\lambda$1199.55 and
$\lambda$1200.22 $N_{\rm a}(v)$ profiles indicates that these lines
are also affected by saturation, even in the weakest component at the
OA velocity.  Moreover, because the \ion{N}{1} lines are quite noisy,
it is difficult to accurately assess and correct for the saturation,
so we conservatively treat the \ion{N}{1} measurement as a lower
limit.  The available \ion{Si}{2} transitions are also strong and
saturated.  Fortunately, the Outer Arm is detected in the \ion{S}{2}
$\lambda \lambda$1253.81, 1259.52 transitions toward HS0624+6907 (see
Figure~\ref{hs0624vel}), and the \ion{S}{2} $N_{\rm a}(v)$ profiles
are in good agreement, so we can obtain a reliable measurement of the
metallicity of the OA toward this QSO.\footnote{Absorption at the OA
  velocity is also apparent in the weakest S~\textsc{ii} transition at
  1250.58 \AA , but the line detection is quite marginal, so we only
  use the two stronger S~\textsc{ii} lines for the measurements.}  Sulfur is
advantageous because it is thought that it does not deplete strongly
onto dust and thus provides a measurement of the overall (gas-phase)
metallicity (Savage \& Sembach 1996, but see the caveats noted by
Jenkins 2009).  The \ion{Fe}{2} $\lambda$1608.45 line should also be
weak enough to yield a good column density, and in this case we can
employ a set of weaker \ion{Fe}{2} lines from the {\it FUSE} LiF
spectra to ensure that the column is reliably measured.  Since iron is
highly prone to dust depletion, this provides insight on the presence
of dust in the OA.  The \ion{Al}{2} $\lambda$1670.79 line is the only
\ion{Al}{2} transition available and is strong enough to be confused
by saturation, but in principle this line can still be used to
corroborate the presence/absence of dust indicated by the \ion{Fe}{2}
measurement.  As we can see from Figure~\ref{hs0624vel}, the
\ion{Si}{4} and \ion{C}{4} doublets are clearly detected in the OA
toward HS0624+6907.  The \ion{Si}{4} $N_{\rm a}(v)$ profiles are in
excellent agreement, and thus the \ion{Si}{4} lines are not saturated.
While the \ion{C}{4} $\lambda$1550.78 transition is relatively noisy,
the \ion{C}{4} $N_{\rm a}(v)$ profiles also appear to be in
satisfactory agreement and are unlikely to be badly saturated.

\section{Distance of the Outer Arm and Complex G}
\label{distance_section}

The kinematical similarity of the Outer Arm, Complex C, Complex G, and
Complex H evident in Figure~\ref{outerarm} is intriguing.  Since
Complex C has been shown to be relatively nearby (Wakker et al. 2007;
Thom et al. 2008), and the OA is not that much farther (if it is
farther at all), it is possible to test whether these clouds are
related by constraining their distances.  For this reason, we have
searched the STIS archive for stellar spectra with implications
regarding the distances of these HVCs.

\subsection{The Outer Arm}

As discussed above, the new constraint on the distance to the Outer
Arm from Lehner \& Howk (2010) raises a question about the nature of
this gas cloud.  However, this constraint is derived from a single
sight line through a highly extended object (see
Figures~\ref{outerarm} - \ref{zoommap}).  As we commented above, the
kinematic similarity and spatial proximity (projected on the sky) of
the OA and Complex C (and other outer-galaxy HVCs) suggests a possible
connection.  Nevertheless, these objects could be at different radial
distances and thus could still be unrelated.  However, Complex C has
been shown to be at a Heliocentric distance of $\approx 10.5$ kpc
(Wakker et al. 2007; Thom et al. 2008), or a Galactocentric radius of
$\approx 12.3 - 13.7$ kpc for the three directions in which is has
been detected toward stars.  Therefore, the Outer Arm and Complex C
have similar radial distances and their three-dimensional locations
are also consistent with a common origin.

As shown in Figures~\ref{outerarm} $-$ \ref{zoommap}, the OA has a
large angular extent, so it is not difficult to find bright stars at a
variety of distances in its direction. To more tightly constain the
location and nature of the OA, in this section we present a minisurvey
for OA absorption toward stars.  There are many high-quality
ultraviolet spectra of stars in its general direction in the {\it HST}
archive.  We have selected a set of stars from the {\it HST} archive
that have been observed with one of the STIS echelle modes at high-S/N
ratio.  The stellar sight lines that we selected for this search are
indicated on the map of the OA in Figure~\ref{zoommap} and are listed
in Table~\ref{distance_tab} with their Galactic coordinates,
Heliocentric and Galactocentric distances, $z$ heights, and the source
of the distance information.  The continuum-normalized absorption
profiles of strong interstellar lines of \ion{C}{2} and \ion{Si}{2}
from the spectra of these stars are plotted in Figure~\ref{starlimits}
along with one of the interstellar lines from the H1821+643 and
HS0624+6907 spectra for comparison.

We show both the \ion{C}{2} and \ion{Si}{2} lines in
Figure~\ref{starlimits} because while the \ion{C}{2} lines are
advantageous because they are the strongest low-ionization metal lines
and thus are the most sensitive probes of low-density clouds, they are
also disadvantageous because they often saturate strongly and
consequently completely hide the sight line component structure.  This
component structure can be recognized in the weaker \ion{Si}{2} lines
while still retaining good sensitivity to low-column clouds.  Several
examples of these advantages and disadvantages are apparent from
comparisons of the \ion{C}{2} and \ion{Si}{2} profiles in
Figure~\ref{starlimits}.  Measurements of (and upper limits on)
$N$(\ion{C}{2}) and $N$(\ion{Si}{2}) derived from the stellar data are
listed in Table~\ref{distance_tab}.  As expected, the \ion{C}{2}
limits are more stringent; comparing the column-density measurements
and limits in Table~\ref{distance_tab} with Galactic ISM measurements
in various contexts (see, e.g., Tripp et al. 2002, their \S 4 and
Appendix), we see that the \ion{C}{2} limits are often well below the
\ion{C}{2} column densities typically detected in the disk of the
Milky Way, and the \ion{Si}{2} limits are less constraining.

Several interesting results are evident from the information in
Figure~\ref{starlimits} and Table~\ref{distance_tab}:  

First, the Outer Arm is clearly detected in absorption toward one
star, HD43818. As shown in Figure~\ref{hd43818detect}, C~\textsc{ii}
and Si~\textsc{iii} absorption is nicely detected and well-aligned
with the OA 21cm emission in this direction.  Two UV absorption
components are evident near the OA velocity; the column densities,
$b-$values, and centroids of these components, measured by
Voigt-profile fitting as discussed above, are listed in
Table~\ref{tab:oa_meas}.  The velocities ($\mid v \mid$) of the UV
absorption and 21cm emission are somewhat lower than the $\mid v \mid$
values in other parts of the OA, but the OA is known to have a
velocity gradient with decreasing $\mid v \mid$ values in this region;
the velocities are consistent with an origin in the Outer Arm.  Of the
nine stellar sight lines that we have examined toward the OA, the
HD43818 sight line is the second-most distant target in both
Heliocentric distance and Galactocentric radius (see
Table~\ref{distance_tab}), so it is perhaps not surprising that the OA
is detected toward this target but not the others.  This indicates
that the OA is beyond the closer stars but in front of HD43818, i.e.,
in the Galactocentric radius range of 9 $< R_{\rm G} \leq$ 10.4 kpc.
While the absence of UV absorption toward HD40893 would seem to be
inconsistent with this result since HD40893 is at $R_{\rm G} = 11.6$
kpc, we note that the 21 cm emission is substantially weaker toward
HD40893 (compare Figures~\ref{highl_1} and \ref{highl_2}), and the
lower column density toward HD40893 could cause the UV absorption to
slip below the detection threshold.  Moreover, the sight lines to
HD40893 and HD43818 are relatively close on the sky, and the
significant differences in the LAB 21cm profiles towards these stars
indicates that the OA is a clumpy structure.  Considering the beam
size of the LAB data (see Kalberla et al. 2005), it is possible that
the HD40893 sight line pierces a very low-density region of the OA.
For this reason, it would be valuable to obtain follow-up observations
of additional distant stars in the general direction of the
OA. Oegerle et al. (2000) have searched for OA absorption in {\it
  FUSE} spectra of the central star of the planetary nebula K1-16, and
the nondetection of the OA toward K1-16 places a similar constraint,
$R_{\rm G}({\rm OA}) > 8.7$ kpc.\footnote{Oegerle et al. (2000) report
  a Heliocentric distance of 1.6 kpc to K1-16.  We note that Cahn et
  al. (1992) determined a somewhat lower Heliocentric distance of 1.0
  kpc for K1-16. Use of the Cahn et al. distance reduces the
  Galactocentric distance constraint to $R_{\rm G}({\rm OA}) > 8.6$
  kpc. This change is very small and has no impact on our discussion.}
K1-16 is particularly useful because this sight line is only 85'' from
the H1821+643 sight line, and it also probes the OA at a greater $z$
height than most of the other stellar sight lines.  

Second, considering the intermediate-velocity components at negative
velocities seen toward H1821+643, we clearly detect these clouds in
the spectra of HD207198, HD208440, and HD209339 (this is most easily
seen in comparison of the \ion{Si}{2} profiles in
Figure~\ref{starlimits}). The velocity centroids and relative line
strengths of these features in the stellar sight lines are quite
similar, but not identical, to those of the H1821+643
intermediate-velocity lines.  The variations of the
intermediate-velocity lines from sight line to sight line are not too
surprising given the angular separations between the stars and the
QSO.  These variations could provide interesting constraints on the
nature of these clouds, particularly for the stellar sight lines,
which are relatively close in the sky.  The velocities of these
features suggest that they are affiliated with the Perseus spiral arm,
and in principle, these sight lines can be used to probe gas flows
affiliated with the Perseus arm. These intermediate-velocity lines are
tangential to this paper, so we defer further analysis of the
intermediate-velocity clouds to a future study.

\subsection{Complex G}

\begin{figure}
\epsscale{1.3}
\plotone{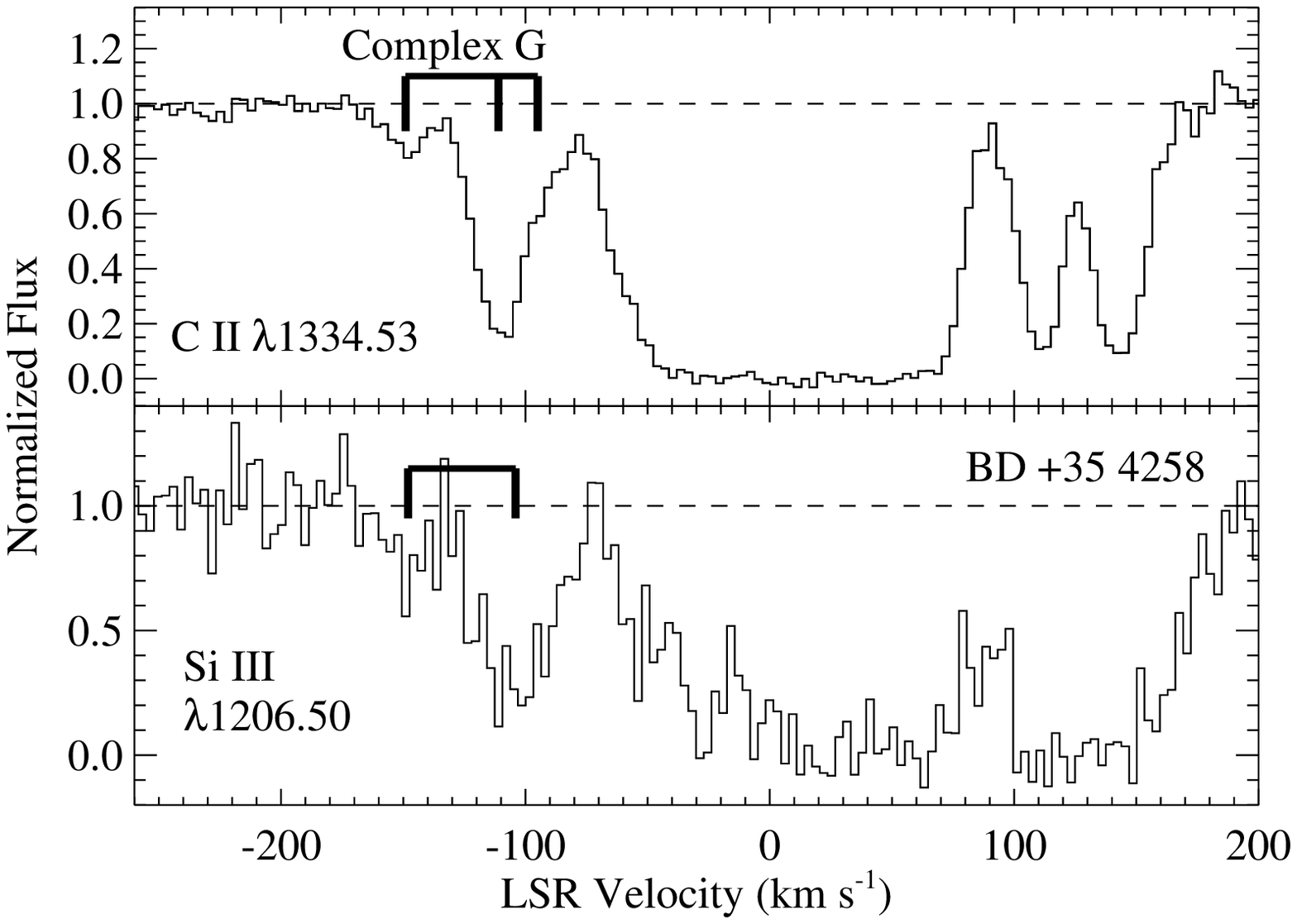}
\caption{Continuum-normalized STIS absorption profiles of the
  interstellar lines of {\sc C ii} $\lambda$1334.53 (upper panel) and
  Si~{\sc iii} $\lambda$1206.50 (lower panel) observed toward the B0.5
  Vn star BD+35 4258, recorded with the E140M echelle mode.  The
  multicomponent absorption lines in the velocity range of HVC Complex
  G are marked with thick vertical tick marks.\label{fig_bd35_compG}}
\end{figure}

\begin{figure}
\epsscale{1.3}
\plotone{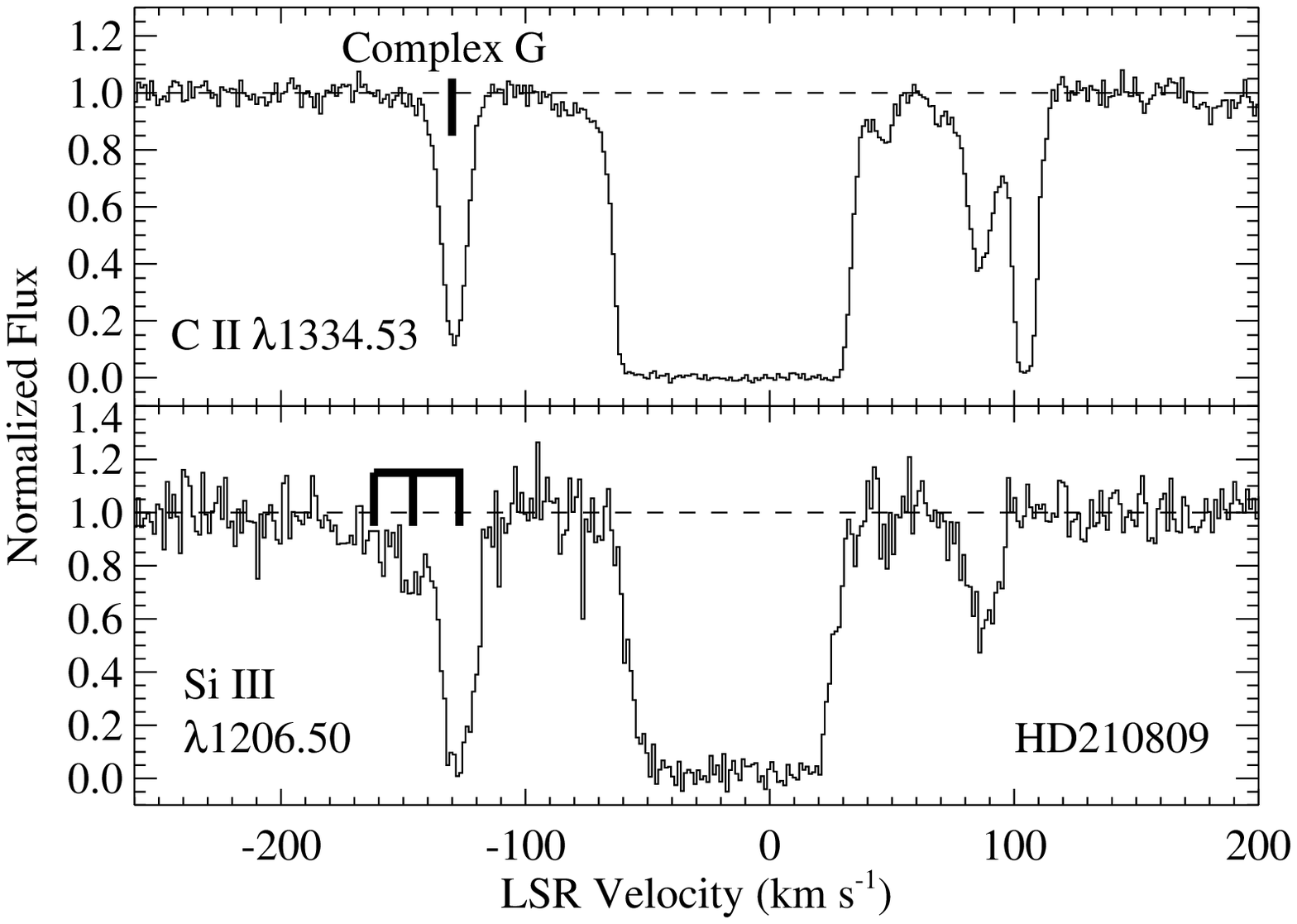}
\caption{Continuum-normalized absorption profiles of the interstellar
  lines of {\sc C ii} $\lambda$1334.53 (upper panel) and Si~{\sc iii}
  $\lambda$1206.50 (lower panel) observed toward the O9 Iab star
  HD210809 with STIS in the E140H echelle mode (see
  Table~\ref{star_obs_log}).  Highly significant absorption lines at
  $v \approx -129$ km s$^{-1}$ (i.e., in the velocity range of HVC
  Complex G) are marked with thick tick
  marks. \label{fig_hd210809_complexg}}
\end{figure}

\begin{figure}
\epsscale{1.3}
\plotone{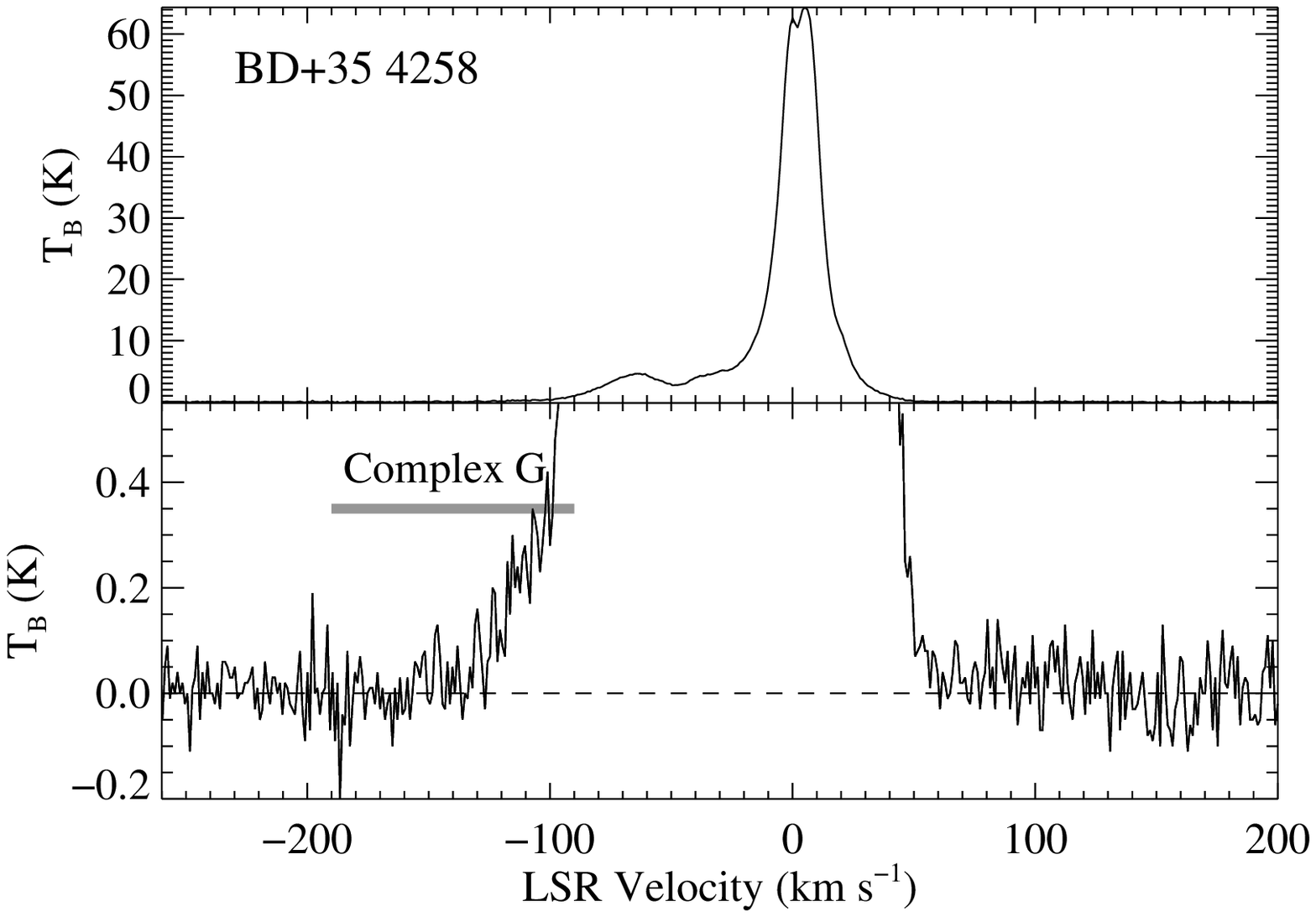}
\caption{{\sc H i} 21 cm emission in the direction of BD+35 4258 from
  the LAB Survey (Kalberla et al. 2005).  Both panels show the
  brightness temperature vs. LSR velocity, but the lower panel is
  zoomed in to more clearly show the 21 cm emission in the velocity
  range of Complex G (indicated with a thick horizontal
  bar).\label{fig_lab_bd35}}
\end{figure}

\begin{figure}
\epsscale{1.3}
\plotone{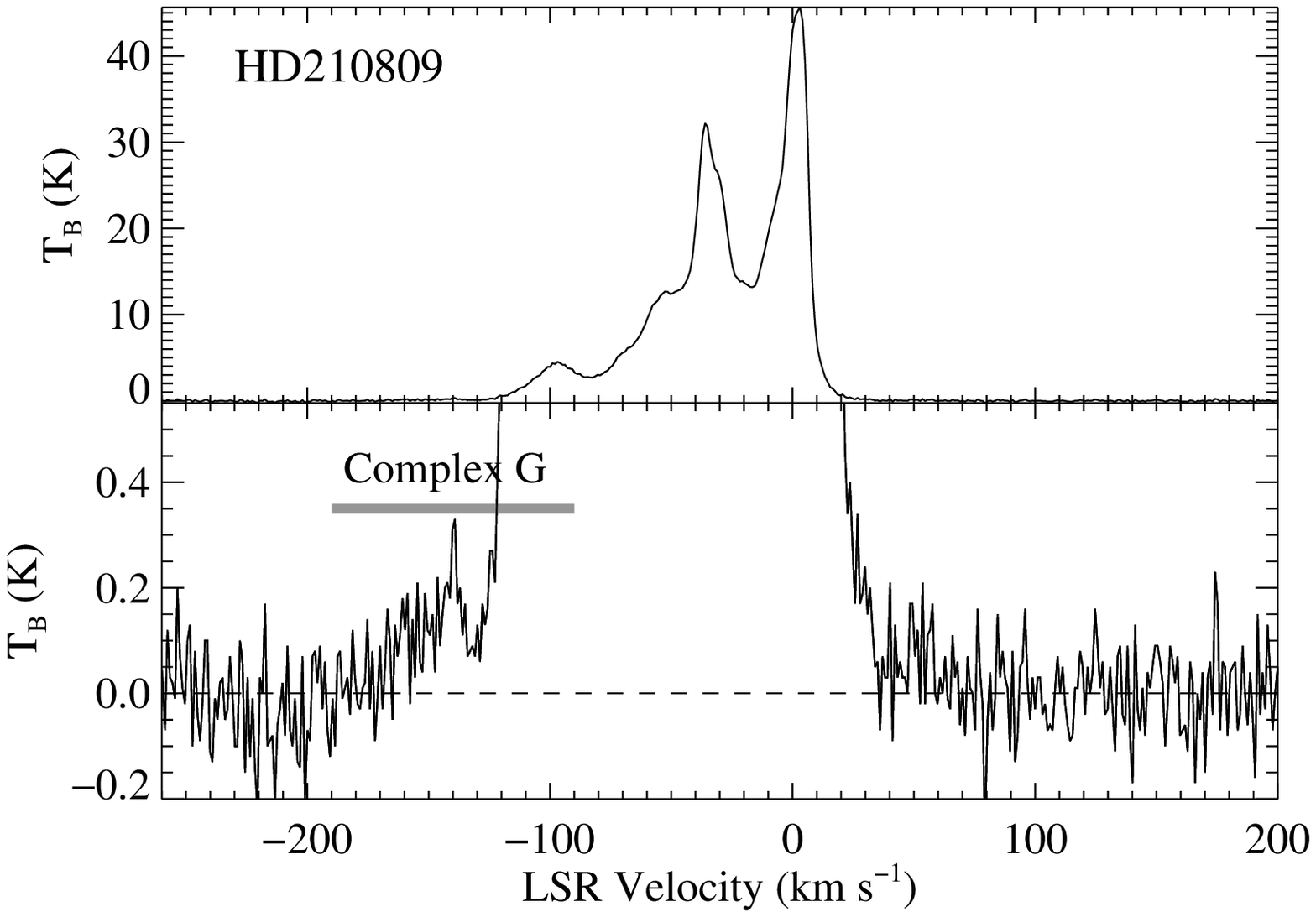}
\caption{{\sc Hi} 21 cm emission in the direction of HD210809 from the
  LAB Survey, as in
  Figure~\ref{fig_lab_bd35}.\label{fig_lab_hd210809}}
\end{figure}

We have found two stars that place upper limits on the distance to
Complex G: BD +35 4258 and HD210809.  The interstellar \ion{C}{2}
$\lambda$1334.53 and \ion{Si}{3} $\lambda$1206.50 absorption profiles
recorded in the STIS echelle spectra of these two stars are shown in
Figures~\ref{fig_bd35_compG} and \ref{fig_hd210809_complexg},
respectively, and the Galactic coordinates and distances of these
stars are listed in Table~\ref{distance_tab}.  In Galactic
coordinates, Complex G extends over $79^{\circ} \lesssim l \lesssim
122^{\circ}$ and $-19^{\circ} \lesssim b \lesssim -1^{\circ}$, and the
velocity range of its 21cm emission is $-190 \lesssim v_{\rm LSR}
\lesssim -90$ km s$^{-1}$ (Kalberla \& Haud 2006).  Both BD +35 4258
and HD210809 are in the direction of this HVC, as shown in
Figure~\ref{zoommap}. From Figures~\ref{fig_bd35_compG} --
\ref{fig_hd210809_complexg}, we see that highly significant \ion{C}{2}
and \ion{Si}{3} absorption lines are detected at Complex G velocities
toward both BD +35 4258 and HD210809, and in many of the profiles,
multiple components are readily apparent.  \ion{Si}{2} absorption is
also detected at the velocity of Complex G toward both stars, but the
\ion{Si}{2} lines are weaker.  The column densities, velocity
centroids, and $b-$values of these \ion{C}{2}, \ion{Si}{2}, and
\ion{Si}{3} lines, determined from the STIS spectra via Voigt-profile
fitting (\S \ref{absmeas}), are presented in
Table~\ref{tab:compg_meas}. The LAB 21cm spectra in the directions of
BD +35 4258 and HD210809 are shown in Figures~\ref{fig_lab_bd35} and
\ref{fig_lab_hd210809}.  Although the 21 cm emission was recorded with
a large ($35.7'$) beam and is blended with lower-velocity emission, we
nevertheless see correspondence between the 21cm emission and the UV
absorption.

\begin{deluxetable*}{llcccc}
\tablewidth{0pc}
\tablecaption{Stellar Sight Lines Toward the Outer Arm: Profile-Fitting Measurements\label{tab:oa_meas}}
\tablehead{Sight line \ \ \ \ \ \ \ \ & Species & Fitted Lines & $v$ (LSR) & b & log [$N$ (cm$^{-2}$)]\
 \\
   \ &          \     & (\AA )       & (km s$^{-1}$) & (km s$^{-1}$) & \ }
\startdata
HD43818\dotfill    & C~II   & 1334.53          & $-79 \pm 1$  & 6$\pm$1          & 13.25$\pm$0.01 \\
 \                 &   \    & \                & $-60 \pm 1$  & 8$\pm$1          & 13.13$\pm$0.04 \\
 \                 & Si~III & 1206.50          & $-77 \pm 1$  & 6$\pm$1          & 12.81$\pm$0.11 \\
 \                 &   \    & \                & $-60 \pm 2$  & 4$^{+5}_{-2}$    & 11.69$\pm$0.23 \\
\enddata
\end{deluxetable*}

\begin{deluxetable*}{llcccc}
\tablewidth{0pc}
\tablecaption{Stellar Sight Lines Toward Complex G: Profile-Fitting Measurements\label{tab:compg_meas}}
\tablehead{Sight line \ \ \ \ \ \ \ \ & Species & Fitted Lines & $v$ (LSR) & b & log [$N$ (cm$^{-2}$)]\
 \\
   \ &          \     & (\AA )       & (km s$^{-1}$) & (km s$^{-1}$) & \ }
\startdata
BD+35 4258\dotfill & C~II   & 1334.53          & $-148 \pm 1$ & 8$\pm$2          & 12.89$\pm$0.05 \\
 \                 & \      & \                & $-110 \pm 1$ & 9$\pm$1          & 13.98$\pm$0.02 \\
 \                 & \      & \                & $-90 \pm 1$  & 6$^{+3}_{-2}$    & 12.92$\pm$0.11 \\ 
 \                 & Si~II  & 1260.42, 1304.37 & $-111\pm$ 1  & 6$\pm 2$         & 12.26$\pm$0.12 \\
 \                 & \      & 1526.71          & $-108\pm$ 5  & 21$^{+27}_{-12}$ & 11.94$\pm$0.29 \\ 
 \                 & Si~III & 1206.50          & $-148 \pm 3$ & 6$^{+12}_{-4}$   & 12.05$\pm$0.16 \\
 \                 & \      & \                & $-104 \pm 1$ & 17$\pm$2         & 12.95$\pm$0.05 \\
HD210809\dotfill & C~II & 1334.53     & $-129 \pm 1$ & 5$\pm$1 & 13.70$\pm$0.01 \\
 \         & Si~II & 1260.42 & $-129 \pm$ 1 & 5$\pm 1$ & 11.82$\pm$0.03 \\ 
 \         & Si~III & 1206.50 & $-162 \pm$ 10 & 16$^{+21}_{-9}$ & 11.83$\pm$0.29 \\ 
 \         & \      & \       & $-146 \pm$ 1  & 4$^{+3}_{-2}$ & 11.74$\pm$0.26 \\
 \         & \      & \       & $-127 \pm$ 1  & 6$\pm 1$ & 12.89$\pm$0.04 \\
\enddata
\end{deluxetable*}

It is very likely that the UV absorption lines in
Table~\ref{tab:compg_meas} are affiliated with Complex G, and this
indicates that this HVC is relatively nearby (see
Table~\ref{distance_tab}).  Given the distance bracket on Complex C
(Thom et al. 2008; Wakker et al. 2007) and the similarity of the
Complex G and Complex C velocities, it is quite possible that these
HVCs are related.  The Outer Arm appears to be located at a similar
Galactocentric radius (see above).  To further test the connections
between these objects, it would be useful to measure the metallicity
of the clouds detected toward BD +35 4258 and HD210809.  The UV column
densities in Table~\ref{tab:compg_meas} are well constrained, but the
\ion{H}{1} column densities must be measured.  Given the low metal
column densities indicated by the STIS data
(Table~\ref{tab:compg_meas}), it might be difficult to detect these
clouds in 21 cm emission, but it might be possible to extract
$N$(\ion{H}{1}) from archival {\it FUSE} data or new COS observations.

\begin{figure}
\epsscale{1.0}
\plotone{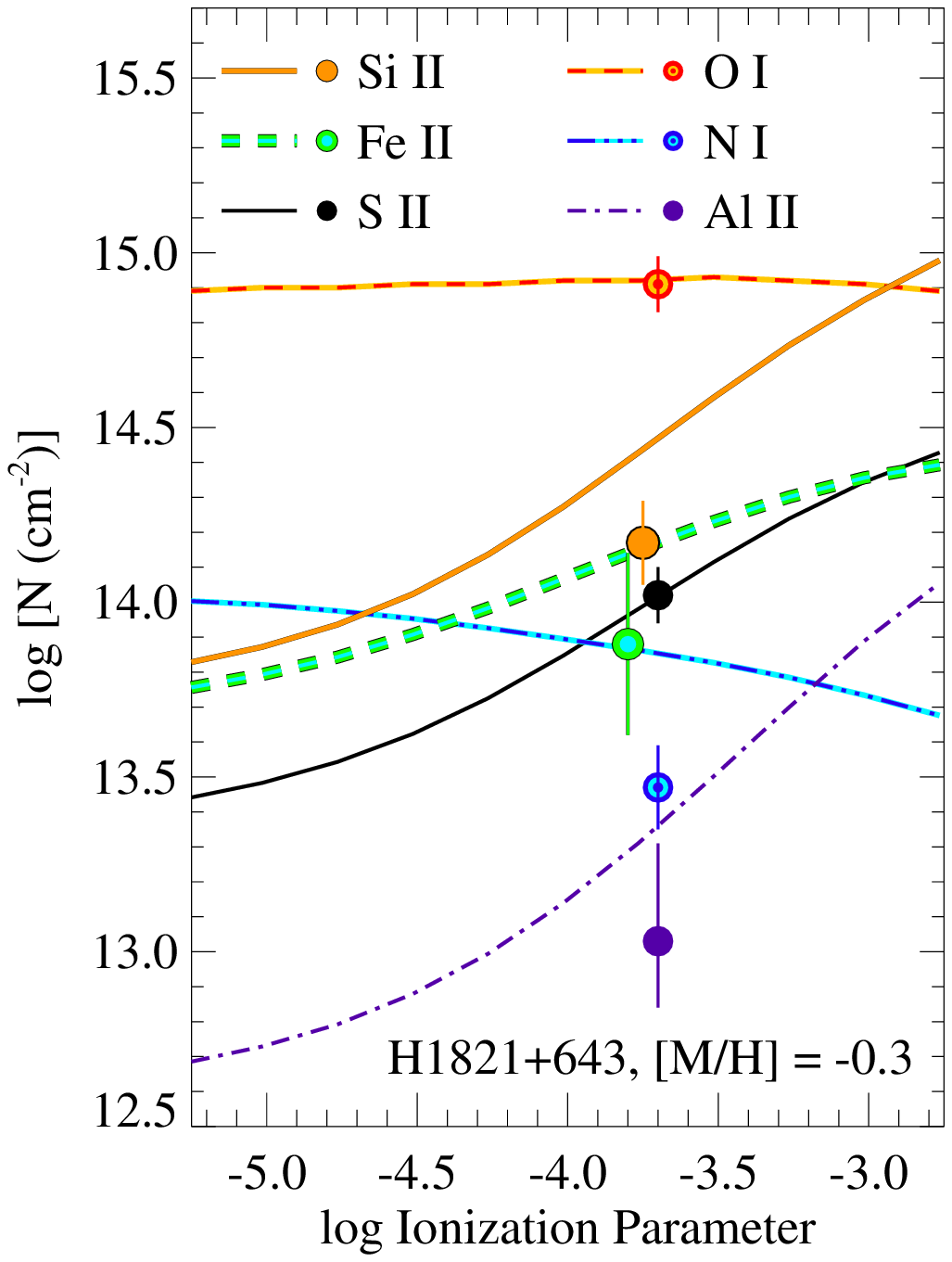}
\caption{Model of gas photoionized by the UV radiation field in the
  ISM of the Milky Way at a $z$ height of $\approx$ 0 kpc (i.e., the
  radiation field in the plane).  The radiation field is based on the
  calculations of Fox et al. (2005, see their Figure 8), and the model
  assumes that the relative abundances follow the solar pattern, as
  determined by Asplund et al. (2009).  Model column densities are
  plotted with various curves (see key at the top of the panel) as a
  function of the ionization parameter $U.$ The Outer Arm column
  densities observed toward H1821+643 are indicated with filled
  circles with $1 \sigma$ error bars, and the different ions are
  identified by the symbol color coding as indicated in the legend at
  the top. The observed column densities are plotted at an ionization
  parameter, log $U = -3.7$, that fits the observed $N$(O~I) and
  $N$(S~II) with an overall logarithmic abundance [M/H] = $-0.30$.
  For a few of the observed points, a slight offset was applied to the
  $U$ position for clarity.  As discussed in the text, this model
  indicates that nitrogen is underabundant and that some dust is
  present in the gas.\label{h1821model}}
\end{figure}

We note that both BD +35 4258 and HD210809 also show high-velocity
absorption lines at {\it positive} velocities, so one might wonder if
the positive and negative high-velocity features are related.  There
is no significant 21 cm emission evident on the positive-velocity side
toward these stars (Figures~\ref{fig_lab_bd35} --
\ref{fig_lab_hd210809}).  Given the difference in the 21 cm emission
and the large velocity separation of the negative- and
positive-velocity absorption lines, it is possible that these
absorption lines arise in unrelated objects in this general direction.
Moreover, a previous study of the nearby star 4 Lac (Bates et
al. 1990), which is in the direction of Complex G, detected this
positive-velocity gas but not the negative-velocity material.  This
indicates that the positive-velocity absorption is closer to
the Sun (Heliocentric $d < 1.3$ kpc), while the negative-velocity gas
is farther away.  However, supernova remnants can produce absorption
lines with such velocity spreads (e.g., Jenkins et al. 1998; Cha \&
Sembach 2000), so the nature of the high-velocity absorption lines
toward BD +35 4258 and HD210809 deserves further investigation.  This
is beyond the scope of this paper, so hereafter we will consider our
Complex G distance constraint with this caveat in mind.
   
\section{Physical Conditions and Abundances}
\label{ionabun}

\subsection{OA Abundances Toward H1821+643}

One of our primary goals in this paper is to measure the gas-phase
metallicity of the Outer Arm based on our absorption-line
measurements.  Toward H1821+643, we are able to measure
$N$(\ion{O}{1}), which is highly advantageous because in
low-ionization gas, \ion{O}{1} is locked to \ion{H}{1} by a strong
resonant charge-exchange reaction (Field \& Steigman 1971). Summing
the columns of the three OA components, and adopting the solar
abundances from Asplund et al. (2009), we find that the OA oxygen
abundance\footnote{We express abundances in the usual logarithmic
  notation, [X/Y] = log (X/Y) - log (X/Y)$_{\odot}$.} toward H1821+643
is [O/H] = $-0.30^{+0.12}_{-0.27}$, including an allowance for the
uncertainty in $N$(\ion{H}{1}) due to the large radio beam (Wakker et
al. 2001).  In this context, oxygen should be only weakly affected by
depletion onto dust (Savage \& Sembach 1996; Jenkins 2009), so this
oxygen abundance is a good representation of the overall gas-phase
metallicity of the Outer Arm.

Abundances of other elements are more difficult to measure toward
H1821+643.  The primary problem is that many species can require
significant ionization corrections, and it is also possible that some
species are depleted onto dust grains.  As we will argue below, the OA
absorption lines likely arise in cool gas.  Therefore the gas is
predominantly photoionized, and we can assess ionization corrections
using photoionization models.  For this purpose, we have the
photoionization code CLOUDY (Ferland et al. 1998) to calculate various
ion column densities as a function of the ionization parameter $U$ (=
ionizing photon density/particle density).  In Figure~\ref{h1821model}
we show the predicted column densities for all of the species that we
have detected toward H1821+643 (Table~\ref{tab:h1821}), assuming that
the gas is photoionized by the ionizing UV flux field in the outer
Milky Way as calculated by Fox et al. (2005) with an
intensity\footnote{The intensity of the ionizing flux at the location
  of the OA absorption is highly uncertain.  However, the
  photoionization models are homologous in the ionization parameter,
  so the ionization corrections are insensitive to this uncertainty. }
of $J_{\nu} = 1 \times 10^{-23} {\rm ergs} \ {\rm s}^{-1} \ {\rm
  cm}^{-2} \ {\rm Hz}^{-1} \ {\rm sr}^{-1} $ at 1 Rydberg.  As
expected, $N$(\ion{O}{1}) is flat over the plotted $U$ range in
Figure~\ref{h1821model} because it tracks \ion{H}{1} precisely as the
gas ionization changes, but the columns of other species, with the
exception of \ion{N}{1}, increase as the gas becomes more ionized
because those ions can remain present in ionized gas that contains
very little \ion{H}{1}.  \ion{N}{1} behaves like \ion{O}{1} and is
coupled to the \ion{H}{1}, but the \ion{N}{1} -- \ion{H}{1} coupling
is weaker.

\begin{figure}
\epsscale{1.3}
\plotone{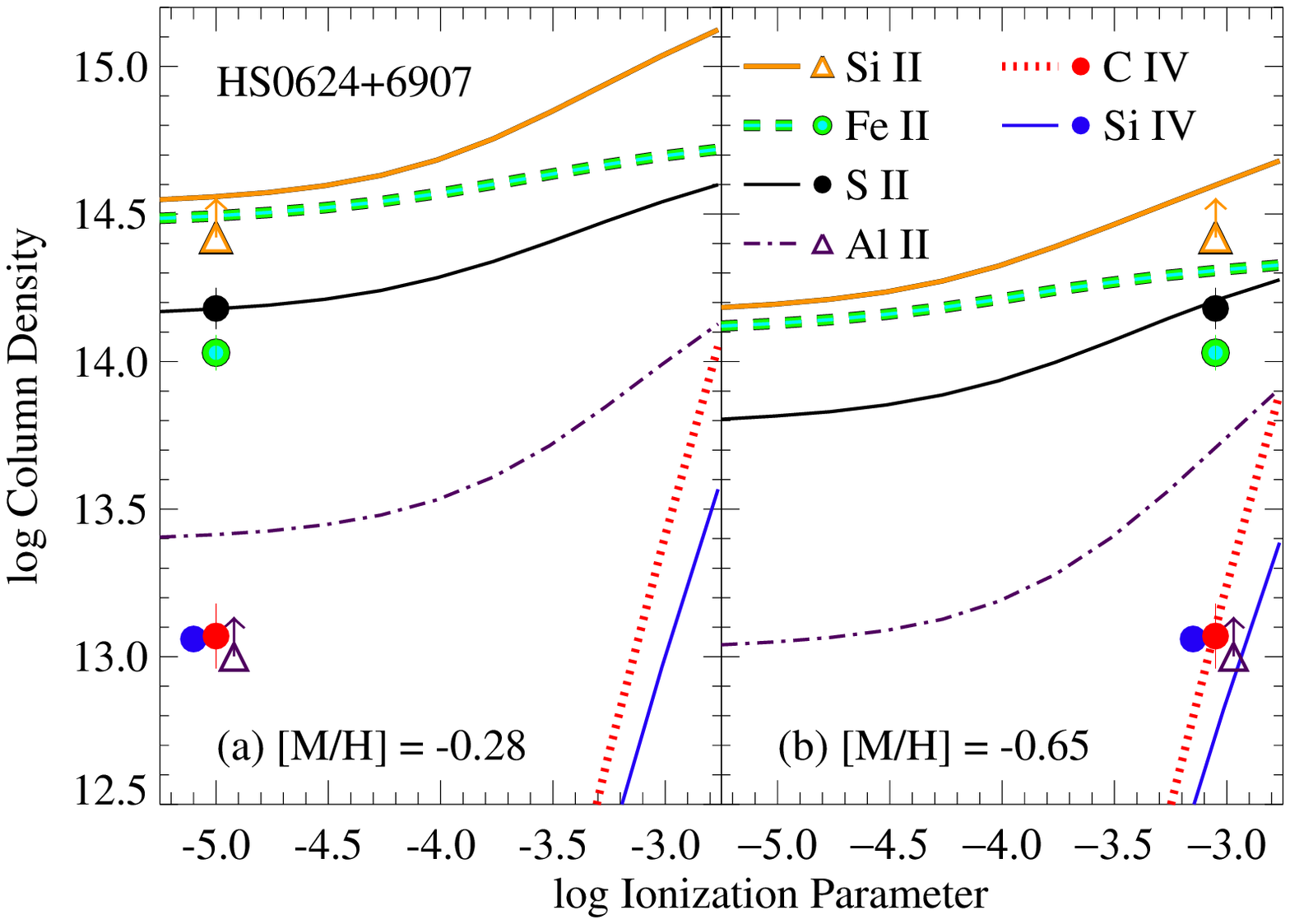}
\caption{Model of gas photoionized by the Galactic UV radiation field
  at $z$ height = 0 kpc (Fox et al 2005), as in
  Figure~\ref{h1821model}, compared to the column densities measured
  in the Outer Arm toward HS0624+6907.  Species that yield reliable
  column densites are plotted with filled circles with $1\sigma$ error
  bars.  Species that have only lower limits on their column densities
  (due to saturation of all available transitions) are indicated with
  triangles with upward-pointing arrows.  The symbol colors and curve
  types identify the species as indicated at the top of panel (b).
  For this sight line, a reliable O~I column could not be obtained, so
  the models are constrained to fit $N$(S~II), and consequently there
  is some uncertainty in the metallicity of the gas due to ionization
  corrections.  Panel (a) shows a model that provides an upper limit on
  the metallicity, [M/H] = $-0.28$ (at values of $U$ lower than the
  plotted range, the low-ion curves are flat and thus indicate the
  same metallicity).  Panel (b) plots the observed column densities at
  the value of the ionization parameter, log $U = -3.1$, that places a
  lower limit on metallicity, [M/H] = $-0.65$.  Increasing the
  ionization parameter decreases the required metallicity, but the
  ionization parameter cannot exceed log $U = -3.1$ because then the
  model would predict C~IV and Si~IV columns that exceed the observed
  values. \label{hsmodel}}
\end{figure}

We can roughly constrain the ionization parameter of the gas by
requiring the model to match the observed columns of the species that
are lightly depleted by dust, \ion{O}{1} and \ion{S}{2}. The
\ion{O}{1}/\ion{S}{2} ratio indicates that log $U \approx -3.7$.  At
this value of $U$, the model $N$(\ion{N}{1}) is 0.5 dex higher than
the observed \ion{N}{1} column.  In fact, the observed \ion{N}{1}
cannot be reconciled with this model at all because this would require
a high value of $U$, and, as can be seen from Figure~\ref{h1821model},
this would cause the model to predict unacceptably high column
densities of \ion{S}{2}, \ion{Al}{2}, and \ion{Fe}{2}; such a model is
ruled out by the measured column densities of these species.  This
high$-U$ model would also produce far more \ion{Si}{4} and \ion{C}{4}
than is observed.  We conclude that nitrogen is underabundant.
Nitrogen underabundance is also observed in the nearby HVC Complex C
(Richter et al. 2001; Tripp et al. 2003; Collins et al. 2007).  We
also notice from Figure~\ref{h1821model} that \ion{Si}{2},
\ion{Al}{2}, and \ion{Fe}{2} are all somewhat overpredicted by the
model at this $U$.  These elements are all prone to depletion by dust,
so this could indicate that there is a small amount of dust in the OA
which removes some of the Si, Al, and Fe from the gas phase and hence
causes the observed columns to be somewhat lower that the model
predictions.  Alternatively, by choosing a lower model ionization
parameter, it is possible to fit the \ion{Si}{2}, \ion{Al}{2}, and
\ion{Fe}{2} columns without requiring any dust depletion, but in this
case the observed $N$(\ion{S}{2}) would significantly exceed the model
\ion{S}{2} column (see Figure~\ref{h1821model}).  Such an excess of
\ion{S}{2} has no obvious explanation; it is more likely that the
preferred model with log $U \approx -3.7$ is correct and there is some
depletion of Si, Al, and Fe.

\subsection{OA Abundances Toward HS0624+6907}

Turning to the HS0624+6907 data, we find that, ironically, the higher
\ion{H}{1} column creates some difficulties because more of the
available metal lines are saturated (\S \ref{hs0624meas}).  In
addition, the HS0624+6907 data are noisier and cover fewer lines.  For
this sight line, we must mainly rely on \ion{S}{2} to estimate the
gas-phase metallicity. This species can exist in ionized as well as
neutral gas, so we must consider the impact of ionization on the
derived metallicity.  Figure~\ref{hsmodel} shows a photoionization
model analogous to the H1821+643 model but adjusted to match the
\ion{H}{1} column observed toward HS0624+6907.  From this figure, we
see that we the ionization parameter is more loosely constrained in
this case.  However, we can exclude large ionization parameters
because if $U$ is too high, the model will predict \ion{Si}{4} and
\ion{C}{4} column densities that exceed the observations.  The right
panel of Figure~\ref{hsmodel} shows the maximum ionization parameter
allowed by this model without exceeding the observed high-ion columns.
The upper limit on the ionization parameter indicates that the
ionization corrections for \ion{S}{2} cannot be large, and the implied
Outer Arm metallicity toward HS0624+6907 is [M/H] $> -0.65$, similar
to the metallicity obtained from the H1821+643 data.  In this case,
the \ion{Si}{2} column is close to the predicted value while
\ion{Fe}{2} and \ion{Al}{2} are somewhat underabundant and, again, may
be modestly depleted by dust.  The upper limit on the metallicity is
derived from the model shown in the left panel of
Figure~\ref{hsmodel}.  In this case, as the ionization parameter is
decreased, the ionization corrections become negligible and the curves
all become flat.  In this case, the \ion{S}{2} column indicates that
the metallicity is [M/H] = $-0.28$, very similar to the H1821+643
metallicity, and the iron and aluminum must be moderately depleted.

\subsection{Physical Conditions}

We noted in \S \ref{h1821measurements} that the H1821+643 absorption
profiles of low ions show clear evidence of three components in the
Outer Arm, and two of these components are required to be narrow. In
addition, while there is clear correspondence between the low ions and
more highly ionized gas traced by \ion{Si}{4} and \ion{C}{4} (see
Figures~\ref{h1821vel}, \ref{h1821nav}, and \ref{fig_multicomp}), the
\ion{Si}{4} and \ion{C}{4} profile shapes are not identical to those
of the low ions.  There are consistent offsets between some of the
low- and high-ion centroids, and the high ions appear to be be broader
in the $v_{\rm LSR} \approx -130$ km s$^{-1}$ component.  These
characteristics suggest that the Outer Arm gas is in a transitional
state.  This gas could be interacting with the ambient halo/disk gas
of the Milky Way, and this interaction is causing the cloud to
fragment and dissipate.  It is particularly interesting that the
\ion{Si}{4} and \ion{C}{4} lines indicate low line widths.  The line
width of the \ion{Si}{4} component at $v_{\rm LSR} = -150$ km s$^{-1}$
indicates\footnote{If the line is predominantly thermally broadened,
  then the gas temperature is directly related to the component
  $b-$value, $T = mb^{2}/2k$.  However, other factors (e.g.,
  turbulence) can contribute to the line broadening, so this
  temperature estimate should be treated as an upper limit.} that the
gas temperature $T < 10^{4.4}$ K, and the \ion{C}{4} in this component
likewise indicates $T < 10^{4.3}$ K.\footnote{The C~IV data are more
  uncertain due to more severe blending with adjacent components, but
  the presence of a narrow component is supported by the consistent
  presence of the sharp edge at $v_{\rm LSR} \approx -150$ km s$^{-1}$
  in both lines of the C~IV doublet.}  These upper limits are well
below the temperatures where these species are expected to exist in
collisional ionization equlibrium (Gnat \& Sternberg 2007).
Interestingly, rather narrow \ion{C}{4} and \ion{Si}{4} components are
frequently identified in the highest resolultion STIS echelle spectra
probing interstellar clouds in and near the disk (Lehner et al. 2011).
Perhaps the gas is simply photoionized, but Lehner et al. (2011) argue
that the narrow and weak \ion{C}{4} and \ion{Si}{4} features do not
arise in regions photoionized by nearby hot stars.  It is interesting
to note that hydrodynamic simulations of a cool gas cloud plunging
through a hot halo can create this type of signature.  For example,
Kwak \& Shelton (2010) have found that the turbulent mixing layers on
the surface of such a cloud can contain a cool phase that is rich in
\ion{C}{4} as well as hotter phases that give rise to species such as
\ion{O}{6}. Toward H1821+643, Sembach et al. (2003) report strong
\ion{O}{6} at $v = -122$ km s$^{-1}$, with log $N$(\ion{O}{6}) =
13.87$\pm$0.18 and $b$(\ion{O}{6}) $\approx$ 22 km s$^{-1}$.  Thus,
this scenario of a cold cloud plunging through, and interacting with,
an ambient medium seems to be consistent with many characteristics of
the OA gas toward H1821+643.  In this situation, it is quite possible
that the gas is not in ionization equilibrium but rather is in an
overionized, relatively cool state. However, due to the strong
blending of the components in the current STIS data, the detailed
parameters of individual components suffer from substantial
uncertainties.  It would be valuable to obtain new STIS observations
of H1821+643 with the E140H grating (which provides substantially
higher spectral resolution) to better contrain the line
widths/temperatures and kinematics of the gas.

\section{Summary}
\label{disc}

As we have summarized above, the extended gas cloud known as the Outer
Arm is usually considered to be part of the warp in the outer Galaxy
and possibly the most distant spiral arm.  However, the recent
observations of Lehner \& Howk (2010) have revealed aspects of the OA
that are not expected in this scenario -- the OA has a high-velocity
component that is inconsistent with Galactic rotation and instead
indicates that the OA kinematics are similar to those of Complex C,
which is close to the OA in velocity and on the sky.  We have
presented additional observations that can be used to further probe
the nature of the OA.  Briefly, we find:
\begin{enumerate}
\item Based on ultraviolet absorption lines, we have measured OA
  abundances in the directions of two QSOs, H1821+643 and
  HS0624+6907. The OA oxygen abundance in the direction of H1821+643
  is [O/H] = $-0.30^{+0.12}_{-0.27}$.  The metallicity derived from
  the HS0624+6907 sight line suggests that the OA could have a range
  of metallicities in different locations with $Z_{\rm OA} = 0.2 - 0.5
  Z_{\odot}$, but the HS0624+6907 metallicity is more uncertain and is
  consistent with the H1821+643 metallicity when uncertainties are
  taken into account.  The OA metallicity is only marginally higher
  than the abundances usually measured in Complex C, $Z_{\rm Comp. C}
  = 0.1 - 0.3 Z_{\odot}$ (e.g., Tripp et al. 2003; Collins et
  al. 2007; Shull et al. 2011).
\item Both the OA and Complex C are underabundant in nitrogen.  This
  is often interpreted to be an indication that the gas is
  ``chemically young'' since nitrogen is synthesized in
  intermediate-mass stars, and thus more time is required to build up
  the nitrogen abundance than is required for species such as oxygen
  that are rapidly produced in Type II supernovae (e.g., Vila Costas
  \& Edmunds 1993; Pettini et al. 1995).
\item High-resolution spectroscopy of several stars in the direction
  of the OA indicates that the object is at a Galactocentric radius of
  $9 - 18$ kpc.  Based on currently available constraints, it is
  possible that the OA and Complex C are at similar distances.
  In addition, we have detected the HVC Complex G, which is close to the
  Outer Arm, in absorption toward two stars.  This places Complex G
  relatively close to the solar Galactocentric radius at $R_{G}$ = $8
  - 10$ kpc.
\item The OA absorption profiles toward H1821+643 show that the OA is
  a complex, multiphase entity with several narrow components,
  including narrow features in the profiles of highly ionized species.
  This suggests that the OA is, at least in part, interacting with the
  ambient gas of the Milky Way. 
\end{enumerate}

This ensemble of information suggests that the Outer Arm might have a
more complicated origin than the usual attribution to the outer warp.
This concept has been proposed before: Davies (1972) suggested a
connection between the OA and Complex C, which he proposed to be
generated by the Large Magellanic Cloud.  More recently, Kawata et
al. (2003) attempted to model the production of both of these
structures by the interaction of a satellite galaxy with the Milky
Way.  While they found that structures similar to the OA + Complex C
could be generated, the absence of the interacting satellite (which
they require to have a mass comparable to the LMC) in the expected
part of the sky poses a problem for this model. At approximately the
same time, the Monoceros Ring structure was discovered (Newberg et
al. 2002), and given the similarity of the kinematics and location of
the Monoceros Ring to the Outer Arm, it is interesting to ask if these
high-velocity gas clouds could be related to the merging satellite
that produced the stellar Monoceros Ring.  We consider this hypothesis
in a future paper.

\acknowledgements

We thank Martin Weinberg and the anonymous referee for many helpful
remarks that significantly improved this paper.  Some of the STIS
data in this paper were obtained through {\it HST} program 9184, with
financial support from NASA grant HST GO-9184.08-A. Additional support
was provided by NASA ADP grant NNX08AJ44G.

{}

\end{document}